\documentclass[a4paper,fleqn,usenatbib]{mnras}

\def\smghy   {{\rm  HS1700.850.1}}

\usepackage{newtxtext,newtxmath}
\usepackage[T1]{fontenc}
\usepackage{ae,aecompl}

\usepackage{graphicx}	% Including figure files
\graphicspath{ {Figures/} }
\usepackage{amsmath}	% Advanced maths commands
\usepackage{listings}

\title[\smghy]{Resolving a merger in a hyper-luminous submillimeter galaxy at z=2.82}
\author[R. W. Perry et al.]{
\parbox[t]{\textwidth}{
R.\,W.\ Perry$^{1}$\thanks{ryan-perry@dal.ca},
S.\,C.\ Chapman,$^{1,2,3}$ %and friends
Ian Smail,$^{4}$
F.\ Bertoldi$^{5}$
}
\vspace*{6pt} \\
$^1$Department of Physics and Atmospheric Science, Dalhousie University, Halifax, NS, B3H 4R2, Canada\\
$^{2}$NRC - Herzberg Astronomy and Astrophysics, 5071 W Saanich Rd, Victoria, BC V9E2E7, Canada\\
$^{3}$Department of Physics and Astronomy, University of British Columbia, 6224 Agricultural Rd, Vancouver, BC V6T1Z1, Canada\\
$^{4}$ Centre for Extragalactic Astronomy, Department of Physics, Durham University, South Road, Durham DH1 3LE\\
$^5$ Argelander-Institute of Astronomy, Bonn University, Auf dem Hugel 71, D-53121 Bonn, Germany\\
}
\usepackage{hyperref}
\date{Accepted XXX. Received YYY; in original form ZZZ}

\pubyear{2018}

\begin{document}
\label{firstpage}
\pagerange{\pageref{firstpage}--\pageref{lastpage}}
\maketitle

% Abstract of the paper
\begin{abstract}
We present the resolved properties of the $z=2.82$ Hyper Luminous Infrared Galaxy (HyLIRG) \smghy, the brightest 850$\mu$m source  found in the SCUBA-2 followup to the Keck Baryonic Structure Survey fields (S$_{\rm 850 \mu m}=$19.5 mJy), and amongst the most luminous starbursts known at any redshift. Using the IRAM-NOEMA interferometer in the highest resolution A-configuration, we resolve the source into two components separated by $\sim$8\,kpc, visible as  blue shifted  and red shifted $^{12}$CO(5-4) lines, exhibiting the expected kinematic properties of a major merger between two gas-rich  galaxies. %actively merging galaxies. 
The combined merger system is traced over 2.3$''$ or 18.4\,kpc. Each component of the merger shows ordered gas motions suggestive of a massive, turbulent disk. %In the $^{12}$CO(5--4) transition, the velocity curves of both components appear to flatten, suggesting the dark matter haloes are beginning to dominate the mass profiles. \textcolor{red}{Chang / take this out?}
We measure the dynamical masses of the blue and red disks as (1.5 $\pm$ 0.2) $\times10^{11}$ M$_\odot$ and (0.71 $\pm$ 0.22) $\times10^{11}$ M$_\odot$ respectively. %Both the NOEMA 2mm and SMA 870$\mu$m continuum maps reveal that the dust continuum flux ratio of the two sources is skewed towards the more gas rich `blue' galaxy component.  
%(S$_{870}$$\sim$15\,mJy in the extended configuration map which spatially resolves the blue component, well fit by a $T_d \sim 29$ K template SED). We therefore deduce that the `red' component must have S$_{870}$$\sim4$\,mJy, and as a consequence a hotter dust temperature (T$_d \sim 42$ K).
%
The more massive disk component shows broad wings in the CO line, offset by $\sim$3\,kpc from the disk centroid along the major axis, and extending to velocities $\sim\pm$1000\,km $ \rm{s^{-1}}$ from systemic velocity. We interpret this as either a possible bipolar outflowing component, or 
more likely a warping or tidal structure in the CO disk.
Comparing the properties of \smghy\ to other submillimeter detected galaxies with comparably bright 850$\mu$m luminosities suggests that  %\textcolor{red}{dense environment, (IAN SAYS NOT TRUE, SEE SECITON 4.4)} or 
ongoing gas-rich mergers, or at least a clustered/group environment  %variety of triggers, ISM conditions, and environments can 
lead to these most extreme starburst phases.
%This is the first time a molecular outflow has been detected in a high redshift SMG, and suggests that we have caught this system at an important yet short-lived phase of its evolution. We find no signs of AGN in the blue (or red) disks, through deep X-ray, infrared SED modelling, or optical spectra. While a buried AGN is always a possibility to drive the outflow, this also suggests the SF driving the outflow, and beginning to shut down its own galaxy-wide SF from SNe-driven outflows, QSO-mode feedback may become dominant at a later phase towards coalescence as the galaxy. We argue, by energetic arguments and comparison to simulations, that the merger has likely advanced enough to have significantly elevated the SFRs of the component galaxies, and therefore the enormous far-IR output of the combined system, and that SNe-driven outflows are capable of driving this molecular outflow.
\end{abstract}

% Select between one and six entries from the list of approved keywords.
% Don't make up new ones.

%%%%%%%%%%%%%%%%%%%%%%%%%%%%%%%%%%%%%%%%%%%%%%%%%%

%%%%%%%%%%%%%%%%% BODY OF PAPER %%%%%%%%%%%%%%%%%%

\section{Introduction}
Our knowledge of %evolution of the universe and 
the history of star formation in the Universe has grown dramatically with the discovery of sub-millimeter galaxies  (SMGs -- \citealt{1997ApJ...490L...5S}; %Smail et al. 1997
\citealt{1998Natur.394..248B}; %Barger et al. 1998
\citealt{1998Natur.394..241H}; %Hughes et al.1998
\citealt{2020RSOS....700556H}%Hodge & da Cunha 2020
). The population was first detected using the Submillimetre Common User Bolometer Array (SCUBA) on the James Clerk Maxwell Telescope (JCMT), and many studies since have dissected the SMG population to further understand their role in the evolution of galaxies in the Universe 
(\citealt{2005ApJ...622..772C}; %Chapman et al. 2005
\citealt{2011MNRAS.413..749G}%Gonzalez et al. 2011
).
SMGs are a massive, gas-rich population which have extremely high luminosities (L$_{\rm IR} \sim 10^{12}$ -- $10^{13}$ L$_\odot$), 
and are typically understood to be dust obscured galaxies that are undergoing a high star-formation rate episode (SFR $\sim 10^2 - 10^{3}$ M$_\odot$ yr$^{-1}$; \citealt{2014MNRAS.438.1267S}; %Swinbank et al. (2014)
\citealt{2017MNRAS.469..492M}). %Michalowski et al. 2017. 
%Surveys revealed to us that the highest density of SMGs occurs at redshifts ranging from $z = 1 - 3$ (Chapman et al. (2005)). SMGs are also now thought to be the progenitors of the present day massive elliptical galaxies.
The brighter SMGs (S$_{\rm 850 \mu m}\sim5$ -- $10$\,mJy) are a relatively rare phenomenon, with space densities of 10$^{-5}$ to 10$^{-6}$ Mpc$^{-3}$ (\citealt{2020MNRAS.494.3828D}). Their rarity may be partly because on average, these galaxies have short depletion timescales ($\simeq$200 Myr; e.g., \citealt{2021MNRAS.501.3926B}),  due to their enhanced SFRs.
%are necessarily short-lived  given their available cold gas reservoir. 
The highest volume density of %radio-selected 
SMGs occurs at $z\sim2$--3, indicating that they are coeval with the peak epoch of star formation (\citealt{2003Natur.422..695C, 2005ApJ...622..772C}; %Chapman et al.\ (2003a), (2005);
\citealt{2014ApJ...788..125S}). 
With the advent of ALMA analysis of an unbiased sample, and the detailed multi-wave followup of the ALMA-identified counterparts,  SMGs are  now established as likely progenitors of the present day massive elliptical galaxies
(e.g., \citealt{2020MNRAS.494.3828D}).%Dudzeviciute et al. 2020

At the extreme bright end, SMGs as luminous as S$_{\rm 850 \mu m}\sim20$\,mJy have been identified in wide-field surveys (e.g., \citealt{2006MNRAS.370.1185P, %Pope et al. 2006%
2017MNRAS.465.1789G,%Geach et al. 2017
2019MNRAS.488.1790L,%Lacaille et al. 2019
2020MNRAS.495.3409S}). %Simpson et al. 2020
The implied high dust masses ($\sim1.1\times10^{9}$M$_\odot$; e.g., \citealt{2021MNRAS.501.3926B}) and SFRs ($>1000$\,M$_\odot$ yr$^{-1}$)  together with compact sizes ($\sim 2.4\pm 0.2$kpc; e.g., \citealt{2015ApJ...799...81S}; \citealt{2019MNRAS.490.4956G})
suggest they must be maximally forming stars %\textcolor{red}{
%radiating at or close to their Eddington limit, %INCORRECT? 
(e.g., \citealt{2008ApJ...688...59Y}). %younger08
A key question is therefore what causes these SMGs to be so luminous? There is now direct evidence that some SMGs are likely  starburst-dominated major mergers, through multiple CO gas components and/or disturbed kinematics (e.g., \citealt{2003ApJ...599...92C}; %Chapman et al.\ 2003b;
\citealt{2006ApJ...640..228T}; \citealt{2010MNRAS.405..219B}; \citealt{2010ApJ...724..233E}; \citealt{2011MNRAS.412.1913I, 2013ApJ...772..137I}; \citealt{2011ApJ...739L..31R}). This has been supported by simulations  (e.g., \citealt{2008MNRAS.391..420S}; \citealt{2010MNRAS.401.1613N}; %Narayanan et al.\ 2010; 
\citealt{2011ApJ...743..159H, 2012MNRAS.424..951H};
\citealt{2019MNRAS.488.2440M}). %McAlpine et al. 2019  
%Indeed, multiple CO gas components and/or disturbed kinematics in some SMGs, support such a merger picture (e.g., Tacconi et al.\ 2006; Bothwell et al.\ 2010; Engel et al.\ 2010; Ivison et al.\ 2011, 2013; Riechers et al.\ 2011).
%This would make SMGs the high-redshift analogs of ultraluminous infrared galaxies (ULIRGs) in the local universe.
%Recently, it has been suggested that 
However, other mechanisms that drive extreme SFRs may also power SMGs, such as  cold mode accretion (CMA; e.g., \citealt{2005MNRAS.363....2K}; %Keres et al.\ 2005; 
\citealt{2009Natur.457..451D, 2009ApJ...703..785D}). %where the star formation in massive, high-redshift galaxies is driven by. 
In CMA-driven galaxies, star formation is sustained by smooth infall and accretion of gas-rich material, and as a consequence are constantly forming stars at high rates.
%This process can result in elliptical galaxies because 
The CMA streams that feed star formation can cause the disk to break up into giant clumps if they have a high enough gas fraction and degree of turbulence. The clumps then potentially migrate inward and merge into a spheroid (\citealt{2009Natur.457..451D, 2009ApJ...703..785D}).
\cite{2009Natur.457..451D} advocated that the CMA phenomenon could power SMGs, and the work has been  extended by others (\citealt{2001ApJ...562..605F}; %Fardal et al. 2001;
\citealt{2006ApJ...639..672F}; %Finlator et al. 2006; 
\citealt{2010MNRAS.404.1355D}; %Dave et al.\ 2010;
\citealt{2015Natur.525..496N}) %Narayanan et al.\ 2015) 
whereby SMGs are massive galaxies sitting at the centers of deep potential wells and fed by smooth accretion.
%\textcolor{red}{
\cite{2010ApJ...714.1407C} %Carilli et al.\ (2010) 
and  \cite{2012ApJ...760...11H} %Hodge et al.\ (2012) 
have provided evidence that one of the most luminous SMGs known in the Universe (GN20) may be driven in this manner, although it also resides in a dense protocluster and may have very recently experienced a major merger. %INCORRECT, part of a protocluster?

Observations of the morphology and kinematics of molecular gas in SMGs is likely the most direct route to understanding the physical mechanisms behind their intense star formation. 
\cite{2008ApJ...682..231S} %Shapiro et al.\ (2008) 
derived a framework for interpreting the growing body of resolved kinematics in distant galaxies, advocating ways to distinguish between a gas-rich merger model and a CMA model  through observations of the gas dynamics and distribution. 
A merger would be represented by tidally disturbed gas with a  starburst nucleus, while a well-defined disk with a smooth rotation curve would be indicative of CMA.
%IE setting up our lack of disturbed gas in h1700.
%  This is simply due to the redistribution of angular momentum. 
 The large-scale gravitational torques induced by gas-rich major mergers are efficient at removing and redistributing angular momentum (\citealt{1996ApJ...471..115B}), %Barnes \& Hernquist 1996)
 thereby funnelling the cold molecular gas into the galaxy's center and producing a nuclear starburst. 
Local ULIRGs (e.g., \citealt{1998ApJ...507..615D}; %Downes \& Solomon 1998;
\citealt{1999AJ....117.2632B}) %Bryant \& Scoville 1999
 and even high resolution simulations of SMGs resulting from mergers (e.g., \citealt{2009MNRAS.400.1919N}) %Narayanan et al. 2009
 show gas reservoirs concentrated in the central kpc or two. 
Less extreme, but still massive star forming galaxies at $z>2$ have shown significant evidence for kinematics of large disk galaxies, with star formation sometimes occurring in ring-like structures at large radii (e.g., \citealt{2008ApJ...687...59G}; %Genzel et al. 2008;
\citealt{2009ApJ...706.1364F}), %Forster-Schreiber et al.\ 2009
with disks  broken into multiple giant clumps of $\sim$1 kpc and 10$^9$ M$_\odot$ (\citealt{2006ApJ...645.1062F}; %Forster-Schreiber et al.\ 2006; 
\citealt{2008ApJ...687...59G}). %Genzel et al. 2008.
\cite{2010Natur.463..781T} %Tacconi et al.\ (2010) 
even found evidence for clumpy CO emission extending over these disks.
 \cite{2006ApJ...651..676E} %Elmegreen \& Elmegreen (2006)
 and \cite{2009ApJ...692...12E} %Elmegreen et al.\ (2009)
 emphasized that CMA could naturally generate large, clumpy disk galaxies.

%Given the large amount of dust obscuring the SMGs, the most sensible way into gaining understanding of these galaxies is to look at the emission and absorption lines that are reproduced by the dust in the far-infrared wavelengths. The question then becomes, what causes these SMGs to be so luminous? Engel at al. (2010) argue that SMGs are the product of major mergers. The arguments for this case came from the emission lines of the SMGs, the dynamics of the material within the source and the morphologies of the object. However, when studying the morphologies of SMGs, its important to realize that UV emission is effected greatly by extinction, so it may unreliable. Using high resolution CO maps, Tacconi et al. (2008) were able to observe multiple morphologies along with complicated gas motions. These observations are either evidence of short lived mergers, or very compact rotating disks, which could be the product of late stage mergers. Narayanan et al. also (2009) show through simulations, that the brightest SMGs in the universe  ($S_{850} \gtrsim 15mJy$) are products of major mergers.

\smghy, one of the brightest SMGs in the Universe (\citealt{2015MNRAS.453..951C}) with $S_{870\rm{\mu m}}$= 19.5$\pm$1.9\,mJy as measured by the  Submillimeter Array (SMA), was discovered in the Keck Baryonic Structure Survey (KBSS) fields (e.g., \citealt{2012ApJ...750...67R}.) %Rudie et al.\ 2012
The HS1700+64 field was imaged by SCUBA-2 with the JCMT to the 850$\mu$m confusion limit as part of the submm-wave followup of 15 KBSS fields, covering $\sim$ 0.2 deg$^2$ (\citealt{2019MNRAS.488.1790L}; %Lacaille et al. 2019
\citealt{2019MNRAS.485..753H}). %Hill et al.\ 2019
The HS1700+64 field contains a massive proto-cluster at $z$ = 2.30 (\citealt{2005ApJ...626...44S}), %Steidel et al.\ 2005
and given the proximity of \smghy\ to the cluster centre ($\sim$1$'$ separation), it was initially thought to be a cluster member (\citealt{2015MNRAS.449L..68C}). A followup mm-wave spectral survey revealed the redshift to be $z=2.816$, well behind the protocluster, and in fact inhabiting a relative void in the Lyman-break galaxy (LBG) redshift distribution of this field (\citealt{2015MNRAS.453..951C}).
$^{12}$CO gas properties of \smghy\ exhibit a strong double peaked line profile (\citealt{2015MNRAS.453..951C}), but the authors could not determine whether it was more consistent with a rotating disk or a merging system due to the low spatial resolution of the spectral imaging (beam size $\sim 3.5''\times2.7''$).
One piece of evidence in favour of a merger was the
R$_{53}$ CO line ratio differences between the two line peaks, which would otherwise imply an
inhomogeneous interstellar medium (ISM) within the system if it were a single large disk rather than two merging components, {although we note that the uncertainties on the ratios make this an inconclusive result.
%\textcolor{red}{Ian suggested we  discuss more of what was known about the system from Chapman et al 2015, can you speak to this? "Need to tell readers the whole story and how the parts fit together"} 
In this paper, we study \smghy\ at high spatial resolution using the IRAM-NOEMA interferometer, resolving the double-peaked CO line of the bright source into two components suggestive of a merger.  This paper has  the goal of determining whether the prodigious SFR of the source is consistent with a major merger explanation, characterizing the source in detail (\S~3), and exploring the implications (\S~4). In \S~5 we present our summary and conclusions.

We use cosmological parameters $\Omega_m$ = 0.286, $\Lambda_0$ = 0.714, and H$_0$ = 69.6 km s$^{-1}$ Mpc$^{-1}$ throughout the paper; at $z=2.816$, this corresponds to an angular scale of 7.995 kpc arcsec$^{-1}$. % $\sim$0.5 Mpc arcmin$^{-1}$. 

\section{Observations}
The following sections outline the detailed observations and methods used to gather the data analyzed throughout the rest of the paper.

%The brightest source in this 850$\mu$m survey was HS1700.850.1 with S$_{850\rm{\mu m}}=19.5\pm 1.9 \rm{mJy}$ as measured previously by the SMA (\citealt{2015MNRAS.453..951C}). 

%\textcolor{red}{Ian: "REPEAT OF THE PREVIOUS PARAGRAPH" 
%These observations revealed a massive, gas-rich galaxy, whose strongly double-peaked CO line profile suggested a rotating system or a major merger. To further study this system, we have obtained higher spatial resolution 2mm observations of \smghy\ using the IRAM-NOEMA.}

\subsection{IRAM-NOEMA }
We observed the redshifted CO(5-4) emission line (rest-frame = 576.3 GHz) in A-configuration in HS1700.850.1 using the IRAM-NOEMA tuned to 150.7 GHz . Observations were taken on March 9 and 10, 2015, using the six-antenna sub-array %\textcolor{red}{
with an exposure time of 14,120\,sec and a pointing center: 17:01:17.779, +64:14:37.85, J2000.0. Flux calibration was achieved by observing various calibrators (3C273, 3C345, B0234+285, B1749+096), while the quasar B1300+580 was used as a phase and amplitude calibrator. Data was processed using the most recent version of the GILDAS software. We resampled the cubes in 90 km s$^{-1}$ channels, and imaged them using the GILDAS suite mapping. % adopting natural weighting. 
A beam size of 0.78$''$ $\times$ 0.41$''$ along a position angle 59 degrees East of North was achieved from the combined tracks.
To obtain flux measurements, we deconvolved the visibilities using the CLEAN task. %, Hogbom. 
We combined the visibilities with natural weighting and inverted the data to an image with 0.13$''$ pixels, calibrated in units of Jy beam$^{-1}$.
We analyzed data from both the WIDEX and ASIC correlators, finding integrated line plus continuum detections of 19.7$\sigma$ and 20.6$\sigma$ respectively.\
\smghy\ is clearly spatially resolved well beyond the beam, showing a Full Width Half Maximum (FWHM) extent of 1.3$''\times0.7''$ with a major axis PA of 48 degrees East of North. The %line plus continuum 
integrated one-dimensional spectrum of the total system and the red and blue intensity maps are shown in Figure~1, and the moment maps are shown  in Figure~2 (see \S3.1.)
%\textcolor{red}{(Ian: mention continuum, SNR, Flux, etc)} 
The 151\,GHz continuum sensitivity (in the collapsed  3.8\,GHz band) is 0.061\,mJy (1$\sigma$), detecting the source with a 10.1$\sigma$ peak flux. The combined NOEMA 151\,GHZ A+D config map reaches slightly deeper at 0.05\,mJy (1$\sigma$). \smghy\ is the only source detected signficantly ($>5\sigma$) in the map.

%constructed by integrating pixels within the half-light diameter.
 
%, and applied the corresponding primary beam correction (negligible at the small 3$''$ offset of the HS1700.850.1 position).

\subsection{SMA}

Followup SMA observations were performed over four different tracks: September 26 and 28, 2014 in the extended configuration (achieving a beam size $\sim 0.72''\times0.74''$ in the combined tracks), and  October 21 and 29, 2014 in the compact configuration (beam size $\sim 2.5''\times1.9''$). 
A subset of these observations were previously described in \cite{2015MNRAS.453..951C}. Here we combine additional data obtained, and separately analyze the compact and extended configuration data. 
% was it all ? or do we add EXT?
All tracks were taken in good weather ($\tau_{225\rm{GHz}} <$ 0.07) with a total on-source integration time of approximately 6 hr (Compact) and 11 hr (Extended). The upper sideband (USB) was tuned to 345 GHz, and combined with the lower sideband (LSB) for an effective bandwidth of $\sim$ 4 GHz at 340 GHz, which yielded a final RMS in the combined observations of 0.9 mJy. The combined map still detects only one significant $>5\sigma$ source (\smghy). The pointing centre was the same as that for the NOEMA observations described above. The data was calibrated using the MIR software package (\citealt{1993PASP..105.1482S}), modified for the SMA. Passband calibration was done using 3C 84, 3C 111, and Callisto. The absolute flux scale was set using observations of Callisto and is estimated to be accurate to better than 20\%. Time-dependent complex gain calibration was done using 1858+655 (0.6 Jy, 21.8$^{\circ}$ away) and 1827+390 (1.8 Jy, 37.7$^{\circ}$ away). In the compact track we detect the source at 9.3$\sigma$, S$_{890\rm{\mu m}}$ = 19.5 $\pm$ %\textcolor{red}{
1.9 mJy, %DOES THIS NEED TO BE 0.9?}, 
after CLEANing the image down to 1$\sigma$. The centroid lies at  $\alpha$(J2000) =17:01:17.767 and $\delta$(J2000) = +64:14:38.15. 
In the extended tracks we detect HS1700.850.1 in the synthesized image at $\sim16\sigma$. The calibrated visibilities in the extended tracks were best fit by a single point source with an extracted  flux density of S$_{890\rm{\mu m}}$ = 15.7 $\pm$ 1.1 mJy 
%\textcolor{red}{(Ian: "where has the $\sim$ 4 mJy gone?")} 
at a position of 17:01:17.779 and +64:14:37.85, consistent within $0.2''$ with the centroid of the 2mm continuum measurement from NOEMA.
This is slightly offset to the south-west by 0.4$''$ from the SMA compact track centroid. 
The extended SMA configuration may have partially resolved out fainter, more spatially extended 850$\rm{\mu m}$ continuum emission which is seen in the compact configuration observations.

\subsection{\it{Herschel} SPIRE }
%*** NOT SURE WHAT TO WRITE HERE \\
We obtained  {\it Herschel}-SPIRE observations for this field (Program ID OT2-ymatsuda-1) from the archive, with an integration time of  1.5 hr.  We processed the images with the {\it Herschel} Interactive Processing Environment (HIPE; \citealt{2010ASPC..434..139O}), following the standard data processing and map-making steps with destriping. The SPIRE FWHM is 18.1, 24.9, and 36.6 arcsec at 250, 350, and 500$\mu$m, respectively. 
 The  bright source corresponding to \smghy\ rises well above the confusion limit. 
 %using the same procedures described in Swinbank et al.\ (2014).
 
 \subsection{\it{Hubble Space Telescope} }
The Q1700 field was imaged with the {\it Hubble Space Telescope (HST)}/ACS using the F814W filter in program ID 10581, PI: Shapley (\citealt{2007ApJ...668...23P}). %Peter et al.\ 2007). 
%For a protocluster mean redshift
 %Inordertoincludeasmanyof the spectroscopically confirmed galaxies as possible, we
%%four pointings (Figure 1). 
%Each pointing was imaged over the course of five orbits of nearly equal exposure time,
The total exposure time of 12,520 seconds corresponds to a sensitivity of 29.0 AB magnitude for a 1$\sigma$
surface brightness fluctuation in a 1$''$ aperture, and
a 28.4 AB magnitude depth for a 10$\sigma$ point source in 
a 0.1$''$ radius circular aperture. The Multi Drizzle script (\citealt{2002ApJ...567..657K}) %Koekemoer et al.\ 2002)
was used to clean, sky-subtract, and drizzle the flat-fielded data products from the ACS CALACS software pipeline. The ACS image alignment to the IRAM frame was checked and refined using a nearby compact and well detected Distant Red Galaxy with a robust CO detection \citep{2015MNRAS.449L..68C}.
The 
FWHM of the point-spread function of 0.12$''$ corresponds
to 1\,kpc at $z$=2.8. 
All fluxes computed for \smghy\ are tabulated in Table 1.

\begin{table}
 \flushleft{
  \caption{Multi-wavelength fluxes for \smghy\ (listed in mJy for continuum and Jy km s$^{-1}$ for lines, and AB-magnitude in the optical).}
\begin{tabular}{llcccc}
\hline
 Wavelength &  Red & Blue & Total & Instrument  \\ 
%{}  & {(GHz)} &  {(hr)}  & (mJy) & {} \\ 
\hline

F814W & 26.7  & 25.4  &  25.1  & \it{HST} \\
 250$\mu$m & -- & -- & 29.2 $\pm$ 5.0 & Herschel\\
 350$\mu$m & --   &  -- & 42.6 $\pm$ 5.3& Herschel \\
 450$\mu$m & -- & -- & 45 $\pm$ 6 & SCUBA-2\\
 500$\mu$m & --  	 & -- &  30.0 $\pm$ 6.4& Herschel\\
 %870$\mu$m & 4.0$\pm$1.3 &15.7$\pm$1.0 & 19.5 $\pm$ 2.1 &SMA\\
% 2mm &0.3$\pm$0.1 & 1.3$\pm$0.1& 1.6 $\pm$ 0.1& NOEMA\\
850$\mu$m & -- & -- & 18.2 $\pm$ 1.1 &SCUBA-2\\
870$\mu$m & -- & -- & 19.5 $\pm$ 1.9 &SMA-COM\\
870$\mu$m & -- & 15.7 $\pm$ 1.1 & --  &SMA-EXT\\
 2mm &-- &-- & 1.42 $\pm$ 0.11& NOEMA\\ 
 3mm & -- & -- & 0.25 $\pm$ 0.07& NOEMA \\
$I$\,CO(5--4) & 1.43$\pm$0.18 & 1.88$\pm$0.20 & 3.25$\pm$0.17 &NOEMA \\
\hline
\hline
\end{tabular}
} 
\end{table}

\section{Results}

\subsection{Evidence in favour of a merger}%Source Properties}
%1)fig1 describe - CO channel maps and  line fitting,  basic discovery of two apparent disks merging, and HST morphology of each.  (Should fit the size of the HST components and describe — I can do this_

The high resolution NOEMA data cube of \smghy\ was analyzed to assess the source structure. 
An average CO line spectrum was measured for the total system by summing all pixels within the elliptical FWHM ($1.3''\times0.7''$), PA=48 deg, East of North), yielding an optimal signal-to-noise ratio total spectrum.
%over the half-light diameter ellipse. % of the full \smghy\ system.
This 1D spectrum (Figure~1) is fully consistent with the integrated flux and properties tabulated for the lower spatial resolution double-peaked CO(5--4) spectrum in \citet{2015MNRAS.453..951C}.
%\textcolor{red}{Ian has an issue with us saying this is consistent given that we say we are using the half light diameter, and in chapman et al. 2015 you used the full emmission apparently. Unfortunately its been so long I honestly dont remember if we did use the half light diameter, maybe we didn't? Or did you in Chapman (2015)?) 
The flux ratio of the two integrated peaks (1.31$\pm$0.30 -- Table~1) is also in good agreement with the lower spatial resolution D-config data (1.23$\pm$0.15) from \citet{2015MNRAS.453..951C} within errors.
To assess the resolved structure, we first constructed intensity maps of each of the two clear CO velocity peaks (Figure~1), integrating all spectral channels within the FWHM of each of the peaks (FWHM $\sim$ 0.86$''$ and 0.98$''$  for the blue and red components, respectively.)  %These channel maps are shown in Figure~1 and were created by summing all the pixels values that were within the emission line of the galaxies. 
The red and blue line intensity map centroids (defined as the pixel with the highest flux) show a clear spatial offset of 0.73$\pm0.17''$, although the major axis of the $0.8''\times0.4''$ naturally weighted beam has a position angle that is unfortunately aligned towards the separation axis of the red/blue components. This limits the interpretation of these offset components as disk or merger without performing further analysis.

\begin{figure*}
\includegraphics[width=7.5cm]{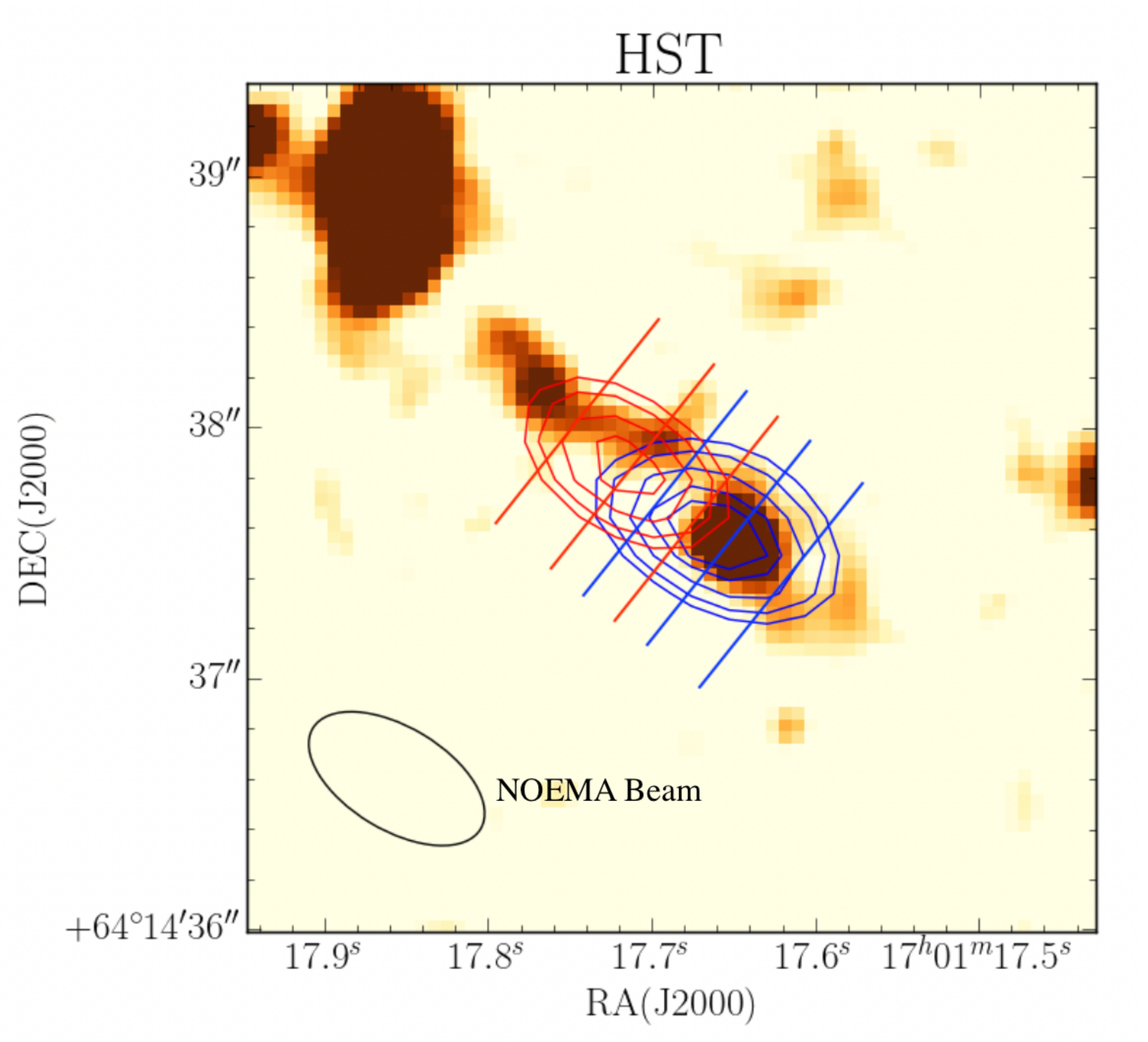}
\includegraphics[width=3.8cm]{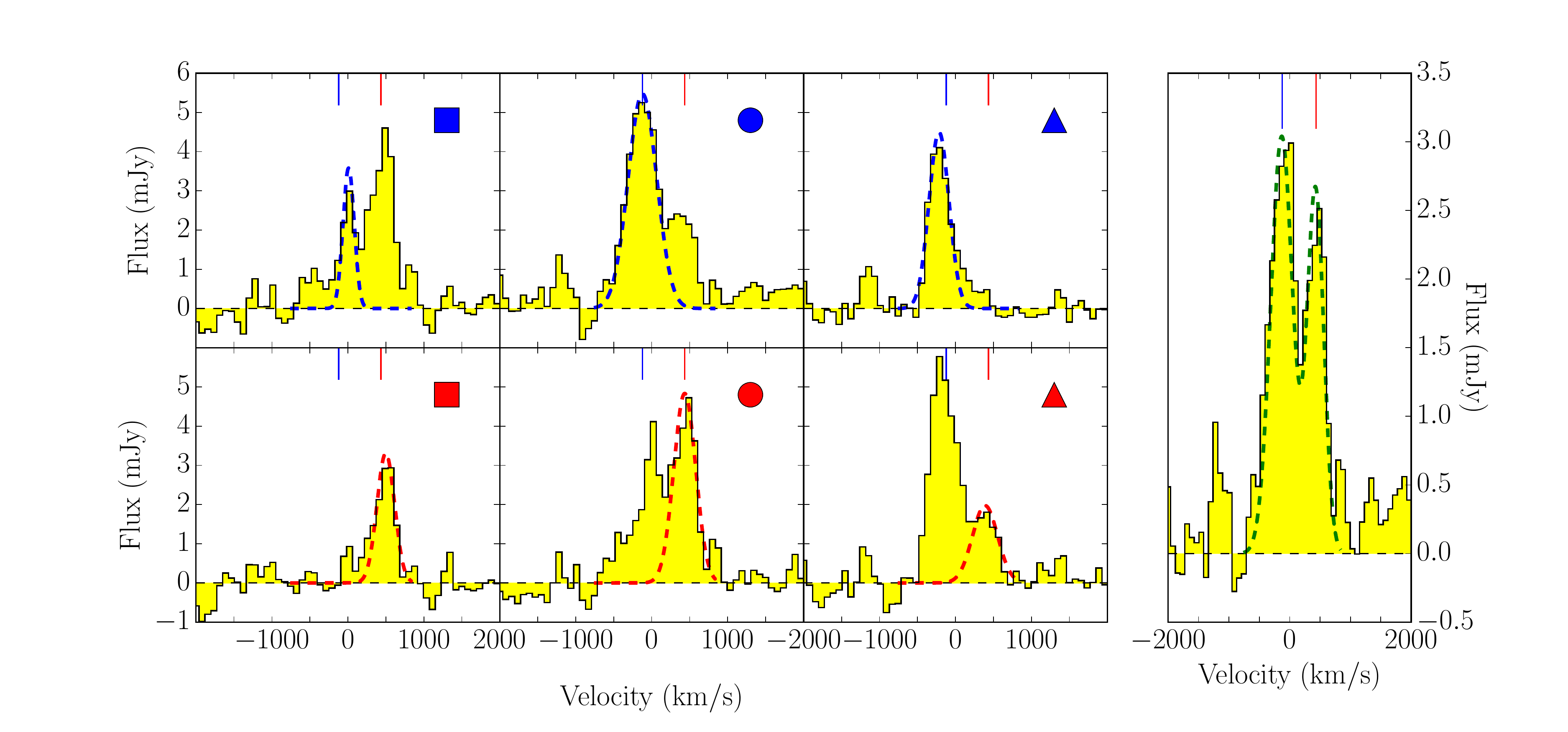}
\hspace*{-0.5cm}  
\includegraphics[width=10cm]{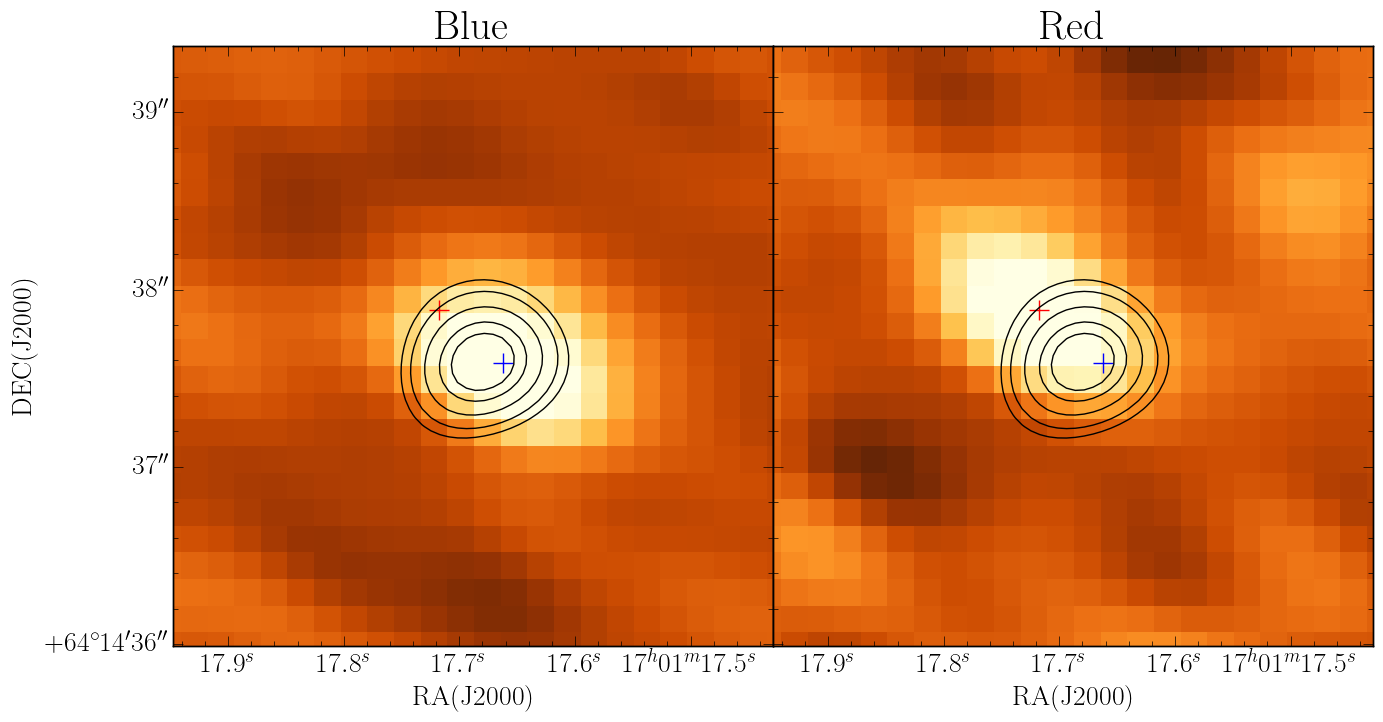}
\caption{ 
 {\bf Top Left:} Contours of the blue and red CO components of \smghy\ overlaid on an {\it HST} F814W image, revealing compact optical emission near the blue component and a clumpy, elongated structure near the red component. The IRAC-NOEMA beam size is displayed in image, which has a position angle that is unfortunately aligned towards the separation axis of the red/blue components. The lines perpendicular to the major axis of the source show the position of the 1D spectra shown in Fig~2. The blue and red contours correspond to 90\%, 80\%, 70\%, 60\% and 50\% of the max flux values for each component.
  {\bf Top Right:} Average continuum subtracted, CO(5--4) one-dimensional spectrum taken over the FWHM ellipse of the full \smghy\ system. The blue and red tick marks at the top of the plot correspond to the components' systemic redshifts, respectively. The zero velocity is defined as the midpoint of the total velocity spread for \smghy.
{\bf Bottom:} Relating the SMA 870$\mu$m  continuum to the CO: intensity maps in CO(5--4) of the \smghy\ blue ({\bf left}) and red  ({\bf right}) components are shown. The contours in the images display continuum maps at $870\mu \rm{m}$ from the SMA in extended configuration. Red and blue crosses indicate the centroid positions of the respective intensity maps. The contours correspond to 90\%, 80\%, 70\%, 60\% and 50\% of the max flux values from the SMA map.}
\end{figure*}

\begin{figure*}
\includegraphics[width=11.7cm]{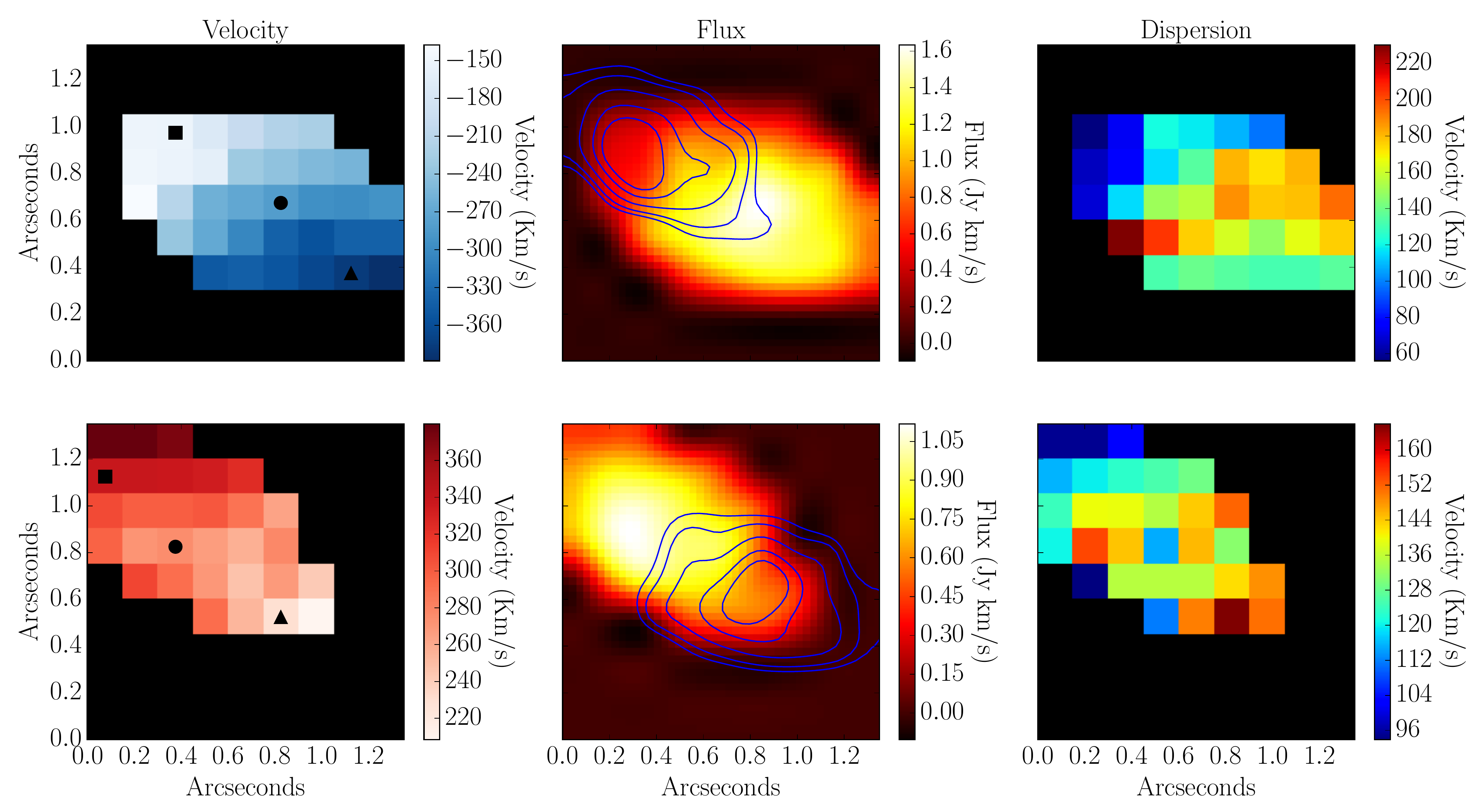}
\hspace*{-1cm}  
\includegraphics[width=11cm]{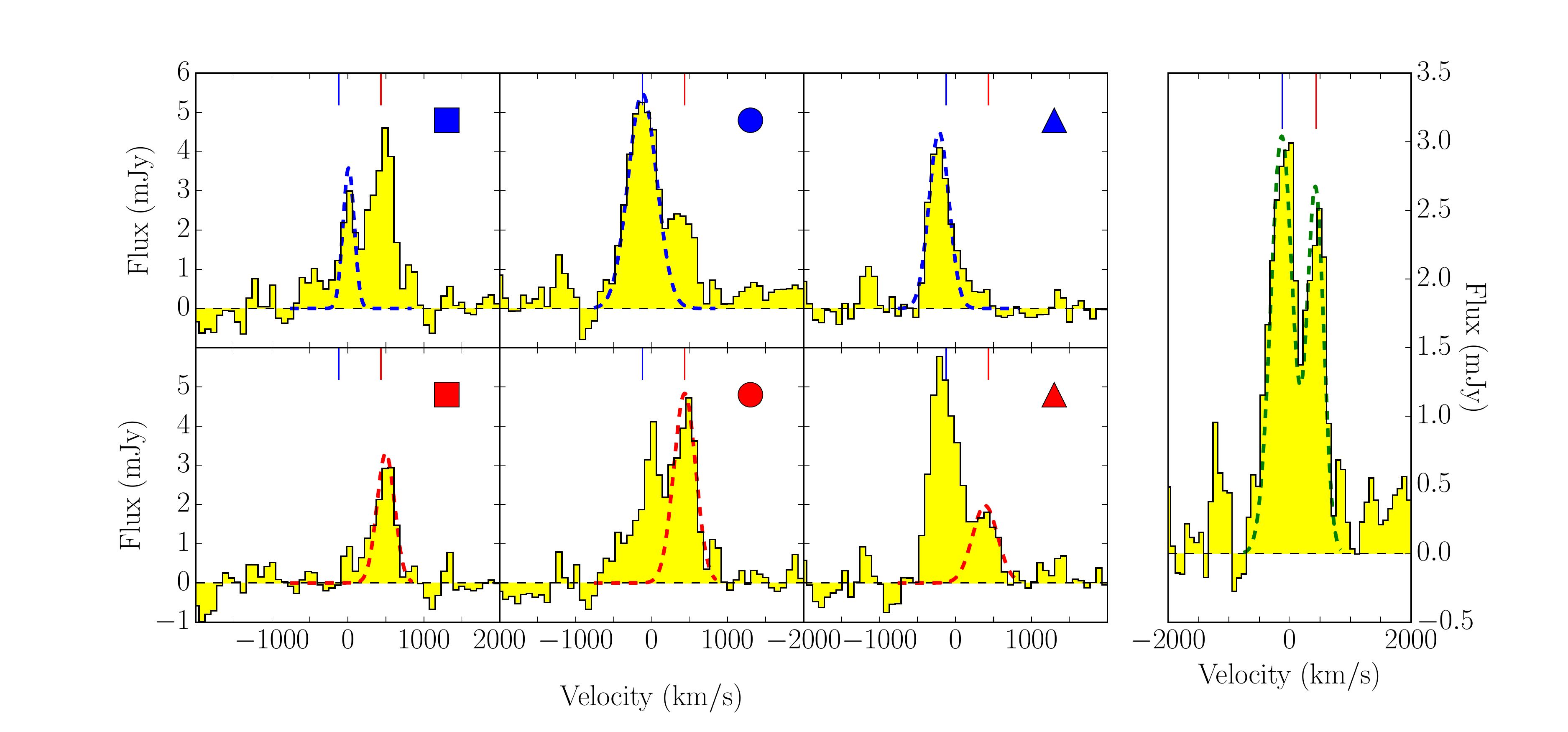}
\caption{ 
{\bf Top:} The velocity, flux and dispersion maps were made from Gaussian line fitting for both the blue (top row) and red (bottom row) components (see text and Appendix B).
%derived from the Gaussian fits to their respective emission lines in each associated pixel (below). 
The positions marked with symbols correspond to the pixel location of the 1D spectra below. The 1D spectra are labelled with the corresponding marker in their plots, whose locations are overlaid in Fig.~1 as well for reference. The flux maps also display the contours of the other respective component for reference. Only pixels with lines detected at $>4\sigma$ are displayed in these maps.
{\bf Bottom:} Representative 1D spectra and Gaussian fits at each of the source components' peak and edges.  The marker displayed in the top right corner of each plot corresponds to the same symbol in the velocity map, showing the pixel location the spectra was extracted from. Fitted Gaussians are overlaid, colour-coded to the red or blue disk component to which they are referring. The blue and red tick marks at the top of the plot correspond to the components systemic redshifts, respectively. 
}
\end{figure*}

\begin{figure}
%\hspace*{-1.5cm}                     
\centering
\includegraphics[width=9cm]{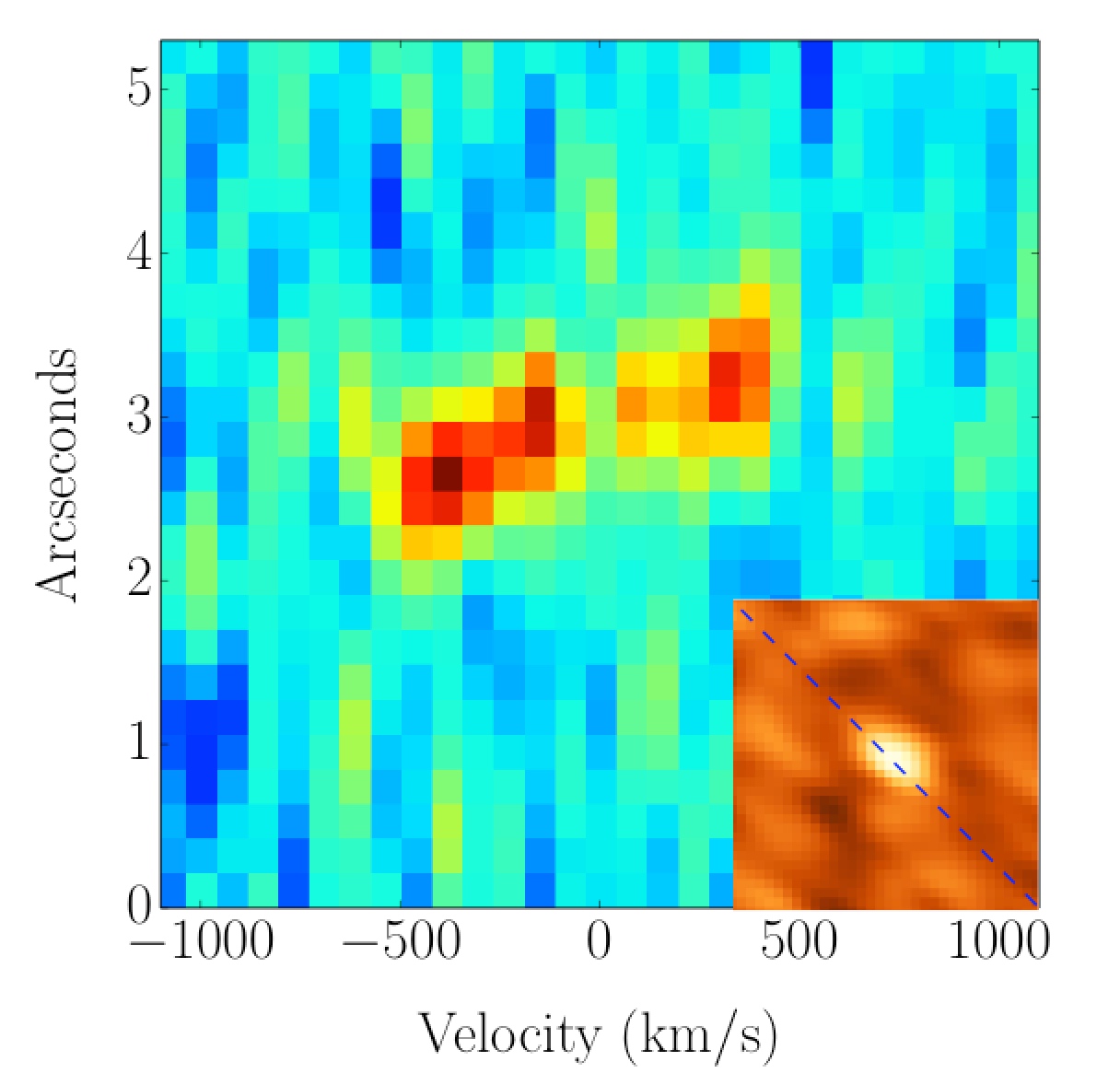}
\includegraphics[width=9cm]{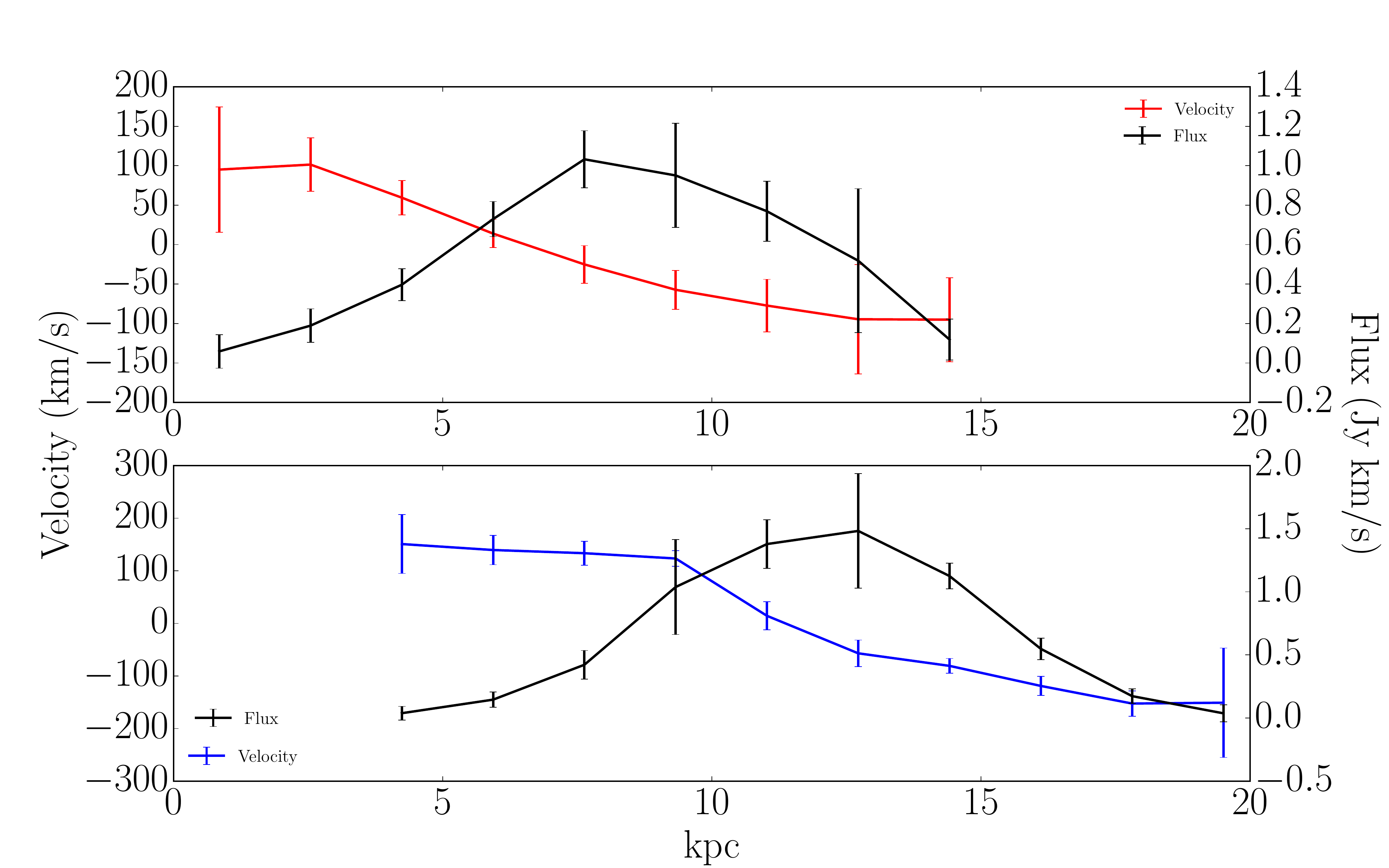}
\caption{
%Top: The combined velocity map for both components.
{\bf Top:} The Position-Velocity (PV) diagram for \smghy,  with an inset showing a 5$''\times5''$ cutout of the collapsed datacube (continuum plus line) and the dashed line showing the slice taken along the major axis of the galaxy in order to obtain the PV diagram. The slices are taken starting from the bottom right (blueshifted) to the top left (redshifted) of the cutout. 
{\bf Bottom:} The rotation curves of each of the components (the blue and red curves, respectively for the blue shifted and redshifted CO(5--4) lines). The black curves show the CO(5--4) intensity profiles. The values were again taken along the disk major axis of the sources. Especially in the blue source, a characteristic disk velocity curve is traced, steepening through the core of the galaxy, and flattening at larger radii.}
\end{figure}

%\begin{figure}
%\includegraphics[width=8.9cm]{SMA_cont.png}
%\includegraphics[width=8.9cm]{SED.pdf}
%\caption{Relating the SMA 850$\mu$m  continuum to the CO: Channel maps in CO(5--4)    of the \smghy\ blue {\bf  Left } and red  {\bf right} components are shown. The contours in the images display continuum maps at $870\mu \rm{m}$ from the SMA in extended configuration. Red and blue crosses indicate the centroid positions of the respective channel maps.
%(compact data: black, extended data: white contours). 
%{\bf  Bottom:} An SED of T$_{\rm{d}}$=36K from the Dale \& Helou (2005) library (green) is fit to the {\it Herschel}-SPIRE, compact configuration SMA, and compact configuration NOEMA measurements, by increasing the temperature of the template until the $\chi^2$ is minimized against the submm photometry. While we only have good constraints on the resolved continuum in the blue galaxy component, we show a possible division of the SED into two components that are consistent with the total SED fit, constraining the red continuum to an inferred SMA 870$\mu$m flux of 3.3\,mJy. The blue (T$_{\rm d}$=28K) and red (T$_{\rm d}$=42k) SEDs correspond to the blue and red CO components.}
%\end{figure}

We next assessed the optical morphology of the NOEMA-resolved structure using {\it HST} imagery in the F814W filter ($\sim I$-band), or rest frame ultraviolet at $z=2.8$. The blue and red NOEMA components are contoured over the {\it HST} image in Fig.~1 revealing a compact source near the blue NOEMA component and a diffuse and clumpy elongated structure near the red component. Their isophotal magnitudes, along with an integrated aperture magnitude of the entire system are reported in Table~1.
There remains some astrometric uncertainty of $\sim0.3''$ between NOEMA and {\it HST} frames, %however this HST morphological result 
although the global astrometry is consistent with the {\it Spitzer} IRAC detection centered on the combined blue/red sources in the low resolution CO detection presented in \cite{2015MNRAS.453..951C}.
The distinct {\it HST} morphologies 
and alignment with the blue/red NOEMA components are consistent with a merger interpretation for this structure, although we caution that clumpy rest-ultraviolet morphologies are not uncommon in SMGs (\citealt{2010MNRAS.405..234S})%Swinbank et al. 2010
, with structured dust explaining the differences between the apparent counterparts of the NOEMA components.
%\textcolor{red}{(Ian: "Not really given that a little structured dust could explain the differences", not sure what Ian is getting at here)}
Figure~1 also shows the 870$\mu$m continuum map (SMA high resolution) overlaid on the blue/red CO intensity maps. Evident here is that the 870$\mu$m emission is weighted more towards the blue CO component, suggesting it may have a larger dust mass (see \S~3.4 for detailed analysis).
%What I have in a previous version of the paper; LOOKED MORE CAREFULLY AND REVISED
%We attempted to align the HST frame to the USNO stars detected in the ACS image, however this procedure leaves an astrometric uncertainty of $\sim$0.3$''$. We then have a further $\sim$0.2$''$ uncertainty in matching the NOEMA frame to the corrected HST image. In the corrected frames  these two faint HST structures appear to lie 0.3$''$ offset to the East from the CO(5--4) lines, and we apply an offset of 0.3$''$ west in Figure~1 such that the brighter knot is aligned with the blue CO line centroid.

Next, we studied the kinematics of the structures through the CO line maps. We prepared the data cube for further analysis by fitting and subtracting the continuum measured in the off-line regions in each 0.13$''$ pixel over %emission spectra within 
the source. We then fit double Gaussians to all spectra measured in the source with integrated line flux detected above 3$\sigma$. %, and
%. Blind, unconstrained fits resulted in sensible line fits only in the $>4\sigma$ regime, and to push the analysis out to larger radius, 
%further constrained the fits to not deviate from  adjacent pixels  by more than 50\,km s$^{-1}$ in peak velocity.
%\textcolor{red}{Ian: "Is this why the velocity field looks so smooth?" 
%Ryan: I dont believe the velocity field had this constraint, at least from what I can see in the data, and if it makes Ian think that the velocity fields are arbitrarily smooth, it probably makes sense to get rid of this sentence?}
%This was justified since the synthesized beam size is $\sim$3$\times$ the pixel size along the beam minor axis, so no large pixel to pixel jumps in velocity should be possible.
%, or by more than 20\,km s$^{-1}$ in velocity dispersion. 
The total system is traced over 2.3$''$ down to this 3$\sigma$ line flux limit, corresponding to  18.4\,kpc. Examples of the continuum-subtracted spectra and Gaussian fits from our source are shown in Figure~2, while the total mosaic of all spectra $>3\sigma$ is shown in the Appendix~ A. From the Gaussian fits, we then constructed velocity, dispersion and line intensity maps, also shown in Figure~2, and we highlighted the locations of the example extracted one-dimensional spectra with specific symbol markers.
The velocity fields for each of the blue and red galaxy components show clear velocity gradients, with an offset in velocity of $\sim$ 350 km s$^{-1}$ between the two components. This further suggests we are seeing a major merger between two gas rich disk galaxies. 
These line maps (Fig~2) also reveal some non-gaussian wings in some of the blue component line extractions (see \S~3.5 for detailed analysis), suggesting again the blue component has distinct properties from the red.

To further characterize the structure, a position-velocity (PV) diagram of \smghy\ was constructed (Figure~3) by slicing through the major axis of the integrated continuum source and plotting the velocity spectrum versus its position within the data. The diagram clearly reveals the two spatially separated sources with a clear break between the two sources where the map drops to the rms noise level, and the offset in velocity of $\sim$ 350 km s$^{-1}$ is obvious. This configuration is not consistent with a single, large filled disk. The rotation fields derived from line fitting  (Figure~2) are discernible in the PV diagrams for each component, with the blue source showing a larger spatial extent over the $\sim$ 500 km s$^{-1}$ span of the velocity gradient. 

Additional evidence for a merger comes from the integrated CO line profiles originally presented in \cite{2015MNRAS.453..951C}.
We revisit this again, combining the new resolved CO(5-4) with the compact configuration observations in the image plane, resulting in similar line fluxes, but with substantially smaller ($\sim$40\%) measurement errors (1.86$\pm$0.14	Jy~km/s and 1.48$\pm$0.13 Jy~km/s for the blue and red peaks respectively.) The $^{12}$CO {\it J}=5 to {\it J}=3 line ratios in the blue and red peaks
then become 0.68$\pm$0.07 and  0.59$\pm$0.08, which now just agree within their 1$\sigma$ uncertainties.

% versus the high-SNR CO(3-2) blended peaks from Chapman+15.
%line	blue	red
%I32 = 	1.26+-0.16	0.88+-0.13
%I54 =	1.84+-0.23	1.50+-0.21
%I54r = 	1.88+-0.20	1.45+-0.18
%I54comb = 1.86+-0.14	1.48+-0.13
%The r53 do just agree within 1sigma uncertainties. r53 = 0.53+-0.07 and  0.60+-0.08

However, these numbers belie the very different blue component line profiles between the $J$=5 and $J$=3 lines.
Considering only the line peak values reveals a significantly stronger difference (30\%) in the $J$=5 and $J$=3 line ratios. However, 
%and that r53 is closer than the relative J=5 and J=3 spectra would suggest because 
the blue component has a much broader $\sigma_V$ in $J$=5 than it does at $J$=3, bringing R$_{53}$ closer to the red component in the integrated lines. 
%Thus we can conclude the gas conditions appear to be significantly different between the Blue and Red components, with caveat that r53 doesn’t end up too different within errors.

%, and confirmed in our high resolution data, which show marginally different $^{12}$CO {\it J}=5 to {\it J}=3 line ratios in the red and blue peaks (1.44$\pm$0.25 versus 1.70$\pm$0.26) respectively.
%\textcolor{red}{(FOUND $0.59 \pm 0.09$ AND $0.69 \pm 0.12$ IN Chapman 2015?)} 
%in $L'{\rm CO}_{\rm \,r53}$ ).

We now understand these two velocity peaks to likely be two distinct merging galaxies, and so these differing line properties (line ratios and line widths) reflect differing conditions in the molecular gas of each galaxy.
It would be hard to understand differing conditions in the opposite sides of a stable disk galaxy.
%The differing ratios of CO r5-3 is further evidence there are two merging galaxies]
%

As a final critical piece of evidence for a merger, we point out that the detailed line properties in the blue and red components are very different. Most prominently, the velocity gradient in the blue component is more than $1.5 \times$ that of the red component (See Table 2), and the blue component shows more stability than the red component under the Toomre Q-parameter analysis (\S~3.3). The blue component also shows broad wings to the CO line, which are not seen in the red component (\S~3.4). If these CO line wings are due to a warping of the disk, they would offer further evidence that we are seeing two merging disk galaxies, with the warping of the wings being a direct consequence of the merger.

We therefore suggest that we have indeed resolved a merger in \smghy, based on eight pieces of circumstantial evidence:
{\it (i)} The red and blue line channel map centroids show a clear spatial offset of $0.73\pm0.17''$; 
{\it (ii)} The {\it HST} optical morphology shows two distinct (and very different) components aligned to the red/blue CO components; 
{\it (iii)} The velocity fields for each of the blue and red
galaxy components show clear velocity gradients, with an offset in velocity of 350 km$ \ \rm{s^{-1}}$ between the two components;
{\it (iv)} A PV diagram reveals the two spatially separated sources with a clean break where the map drops to the rms noise level, inconsistent with a single, large filled disk;
{\it (v)} Different gas conditions in the red and blue peaks ($^{12}$CO $J$=5 to $J$=3 integrated line ratios and line widths); 
%(Chapman et al.\ 2015b);
{\it (vi)}  The velocity gradient in the blue
component is almost twice that of the red component, and
the blue component shows more stability than the red component
under the Toomre Q-parameter analysis (discussed in more detail in \S~3.3);
{\it (vii)} The 870$\mu$m continuum emission appears to be more localized in the blue component, suggesting it has a larger dust mass (discussed further in \S~3.4);
{\it (viii)} The blue component
 shows broad wings to the CO line (Fig~2), which are not seen in the
red component, possibly due to the interaction between the disks (discussed further in \S~3.5).

We analyze the properties of these two merging components in the following sections.

\subsection{Rotation curves and mass measurements}
%2) details of what you did and analysis of PV and rotation curve, how constructed, errors etc - describe fig3 in detail Describe mass estimates of individual disks from rotation curves, compare to dispersion mass of each 'disk' component in prev subsection, and compare to overall mass in chapman+2015 (from single gaussian fit to whole structure and the 'rotation' of two peaks -- this should define a halo mass fig2 describe - PV diagram and rotation curve analysis tell us more about each system, confirm merger of two galaxies, and lead to mass modelling.

From the velocity maps, we constructed one-dimensional rotation curves for both the red and blue components (Figure~3) by slicing through the major axis of each source (defined by the maximum velocity spread) and extracting the peak velocity measurements recorded in each emission spectra. The CO(5--4) intensity profiles are also plotted for each source for reference.  The errors shown are derived from the line fitting uncertainty. 
The CO rotation curves of both components change relatively smoothly and monotonically between velocity extremes, however both show signs of %to be traced out to near the 
flattening towards the edges of each curve. 
This would be expected if we were probing the dark matter halos beginning to dominate the mass profiles, although the result is only significant at $>3\sigma$ in the positive velocities of the blue component (and otherwise is consistent at $\sim2\sigma$ with no flattening).
%\textcolor{red}{This may suggest we have detected rotating gas out to where the dark matter halos begin to dominate the mass profiles. This would be remarkable given the relatively high $J$=5 CO transition probed, since this must be tracing highly star forming gas. (Ian suggests this is not correct and wants to get rid of this in the text)}

We next use these rotation curves to estimate the disk masses. % structures based on their rotations. 
%
%Our calculation of measuring the rotation curves of each SMG allows us to estimate their masses assuming that the emission line is emanating from a rotating disk. 
In this case, the dynamical mass is given by, 
\begin{equation} 
\mathrm{M_{dyn}} <\sin^2(i)> [M_\odot] = 2.35 \times 10^5 V^2 R 
\end{equation}
where $V$ is the line of sight circular velocity in km s$^{-1}$, $R$ is in kpc, and $i$ is the inclination of the galaxy. For our calculations, $V$ has already been calculated as the velocity gradient, and the $R$ used is the Half Width Half Maximum of the intensity profiles in Figure 3. We adopt a mean inclination angle, suitable for a collection of randomly oriented disks of $ <\sin(i)>= \pi / 4 \simeq 0.79$ (see \citealt{2009ApJ...697.2057L}). %Law et al.\ 2009).
Using this rotational estimate, the dynamical mass for the blue and red components of \smghy\ are (1.5 $\pm$ 0.2) $\times10^{11}$ M$_\odot$ and (0.71 $\pm$ 0.22) $\times10^{11}$ M$_\odot$ respectively. It is worth noting that if the inclination of these galaxies is closer to an edge on view (smaller $i$) then the dynamical mass is being underestimated by our calculations. Conversely, if the inclination is more face on (larger $i$) then these estimates are larger than their actual by a factor of $\sim 1.6$.

 We can also compare these individual component dynamical masses with an estimate of the total halo mass, by taking the systemic velocity of each of the disks and applying the virial theorem with the velocity separation of the two components as input (e.g., \citealt{2006MNRAS.371..465S}). %Swinbank et al.\ 2006).
 %A crude estimate for the total mass of the major merger is also calculated using both the total peak to peak velocity of the whole system, as well as a single dispersion fit for both galaxies. 
 Using equation 1, 
% \begin{equation} 
%\mathrm{M_{dyn}} [M_{\odot}] = 1.56 \times 10^{6}\sigma^{2}R 
%\end{equation}
the mass of the total system is 
(16 $\pm$ 2){$\times10^{11}$ M$_\odot$, nearly eight times the summed mass of the two components, although we note this comparison does depend upon the inclination of each galaxy and their orbit. Since this halo mass estimate only includes the line of sight velocity component and their projected separation, the actual halo mass could be higher than this. 
%For simplicity and to easily compare with other SMGs, we will use equation 2 for  subsequent calculations that involve the dynamical mass.
 %, and using equation 2 yields a mass of 
 %1.0$\times10^{12}$.

 \subsection{Stability of \smghy}
To further quantify the structure of this source, we look at Toomre's Q criterion, which characterizes the stability of a gas rich disk  against local axisymmetric perturbations. In order for a galaxy to be stable it must have a $Q$ \textgreater 1, otherwise it will fragment and collapse into giant dense clumps. The Toomre parameter is calculated from  
\begin{equation} 
Q = \frac{\sigma_{r}\kappa}{\pi G \Sigma_{\rm{gas}}} 
\end{equation}
%
%The mass surface deinsity I used was ~ 0.01 (in km)
%
where $\kappa$ is the epicyclic frequency, $\sigma_{r}$ is the line of sight velocity dispersion, and $\Sigma_{\rm{gas}}$ is the mass surface density of the gas (\citealt{1964ApJ...139.1217T}). %Toomre, 1964. 
%\textcolor{red}{
Here we have adopted $\kappa$ = $aV_{\mathrm{max}}/r$ which is appropriate for a uniform disk, where ($a = \sqrt{3}$), and we have assumed that the measured velocity dispersion is equal to $\sigma_{r}$. By estimating $\Sigma_{\rm{gas}}$, we derive $Q = 1.50 \pm 0.20$ and $1.20 \pm 0.18$ for the blue and red disk components respectively. 
%\textcolor{red}{I changed this slightly, as it said that we used the dynamical masses to estimate the gas mass surface density, which isn't true since we just used the gas mass not dynamical}
These are similar to typical Q values for gas rich star forming galaxies at $z$$\sim$2 (\citealt{2014ApJ...785...75G})%Genzel et al.\ 2014)
, and indicate that both disks are still very stable despite the ongoing merger. %We can also improve this picture by looking at the mass ratio of the components.
The dynamical mass ratio estimate for the merging components equates to $0.47 \pm 0.12$. This is consistent with being a major merger (which have mass ratios of 1:3 or closer to unity; eg., \citealt{2010MNRAS.405..219B}; %Bothwell et al. 2010
\citealt{2010ApJ...724..233E}%Engel et al. 2010
). The lack of a highly turbulent component (small measured $\sigma_r$ for the $V_{\mathrm{max}}$) may indicate the galaxies are being found at their first approach.
%The Q values for the blue and red components are displayed in Table 2. 

\subsection{Continuum properties and SFRs}
We next attempt to constrain the rest-frame far infrared continuum properties of the red and blue galaxy components, in order to better estimate their SFRs.
%\textcolor{red}{Ian: explain why? use Herschel for (T$_{\rm d}$=37.1$\pm$3.1)}
The 2mm continuum map from NOEMA does not significantly resolve the two components, due to the position angle of the synthesized beam lying near the merger axis of the components. %, although does demonstrate that the c
%Again, 
Intensity maps of both the blue and red CO components of \smghy \ can be seen in Figure 1 with the extended configuration 870$\mu \rm{m}$ imaging contoured over top. The offset of the 870$\mu \rm{m}$ centroid to the blue component centroid is $0.22'' \pm 0.08"$, compared to that of the red component, $0.51'' \pm 0.12"$. These results were obtained by combining the  beam uncertainties  (as the beam$_{\rm{FWHM}}$/SNR) at both wavelengths in quadrature. 
The 850$\mu \rm{m}$ continuum  is better aligned with %to be almost completely on top of the 
the blue galaxy component, suggesting that this component may carry a majority of the 850$\mu \rm{m}$ flux.  
%makes it difficult to cleanly separate the 2mm continuum, manifesting as an extended continuum source with a  centroid  more consistent with the blue component and SMA source position. We model the continuum emission from the merging pair with the blue component carrying the entire 15.7\,mJy from the high resolution, extended configuration SMA measurement, and treat the red component as being consistent with the $\sim4$\,mJy $3\sigma$ limit of this dataset at the red component centroid. 
%
%For the 2mm continuum, we fit a two beam model to the extended source with positions fixed to the blue and red CO channel map centroids. This results in a blue/red 2mm flux ratio of 3.9, as listed in Table~1.
%then is most likely only carrying the minimum detection level of 4$\sigma$ and thus the blue component would have the remainder of 15$\sigma$. 

%\textcolor{red}{Not sure if you accidentally added this paragraph back in? Thought we got rid of this part}
We next fit a Spectral Energy Distribution (SED) template from the \cite{2005ApJ...633..857D} %Dale \& Helou (2005) 
library  to the integrated continuum measurements of the source using {\it Herschel}-SPIRE, compact configuration SMA, and compact configuration NOEMA measurements. The  temperature of the template is increased until the $\chi^2$ with the photometry is minimized, resulting in an SED of T$_{\rm d}$=36.1$\pm$3.1\,K. This is comparable to the average T$_{\rm d}$ measured by \cite{2010MNRAS.409L..13C} and \cite{2014A&A...561A..86M} for SMGs followed up with SPIRE, as well as \cite{2014MNRAS.438.1267S} for a large sample of field SMGs using ALMA fluxes with deconvolved constraints on the SED peak from SPIRE.  
Given the uncertainty in T$_{\rm d}$, our new estimate for the L$_{\rm IR}$= 4.3$\pm 1.4 \times10^{13}$\,L$_\odot$ is not appreciably different from that estimated previously (without the SPIRE photometry) of 3.8$\times10^{13}$\,L$_\odot$   \citep{2015MNRAS.453..951C}. Based on the scaling relations mentioned in \cite{2021MNRAS.501.3926B}, the dust mass of the total system would then be $\rm M_{\rm dust} = 3.2 \times 10^{9}.$

%While we have better constraints on the resolved continuum in the much brighter blue galaxy component, we show a possible division of the global SED into two components that are consistent with the total SED fit, and the integrated CO luminosities of the two components, 
%assuming a fixed L$_{\rm CO}$ to L$_{\rm IR}$ for both. 
%We constrain the red continuum to the measured 2mm continuum and the inferred SMA 870$\mu$m flux of 4\,mJy as noted above. The best fit SEDs consistent with these constraints have dust temperatures for the blue, T$_{\rm d}$=28\,K and red, T$_{\rm d}$=42\,K.
%SEDs correspond to the blue and red CO components.
While we cannot unambiguously resolve the continuum in the two components, and thus cannot reasonably infer individual L$_{\rm IR}$ from SED modelling, the weighting of the 870$\mu$m emission centroid towards the blue component is consistent with the CO(3-2) luminosity of the blue component being 1.4 $\times$ higher than that of the red component.  A corresponding division of the total SFR (3800$\pm 600$ M$_\odot$ yr$^{-1}$) by the same ratio would yield 2216$\pm 500$ and 1583$\pm 300$ M$_\odot$ yr$^{-1}$ respectively for the blue and red components. %, with a large uncertainty of a factor $\sim$2 on the red component.

%Using the data in table 1, we were able to fit two Spectral Energy Distributions (SEDs) for the red and blue components, as well as the total source. %The SEDs were fit by constraining the 850$\mu m$ flux of the blue component to 15mJy, and the Red component to 4mJy. 
%The integral of the blue curve was also constrained to be 1.5$\times$ the red. This was done due to the FIR luminosity scaling with the radio synchrotron at very long wavelengths. Thus, by imposing these restraints, and adding the blue and red SEDs together, the total SED was determined. From the templates used for the Blue and Red components, we found that their respective gas temperatures are 28.8K and 37.9K. Using the inferred L$_{\rm IR}$ from these SED models, the blue and red components have a corresponding SFR of 2500 and 1300 M$_\odot$ yr$^{-1}$ respectively. 

\begin{figure}
\centering
\includegraphics[width=9cm]{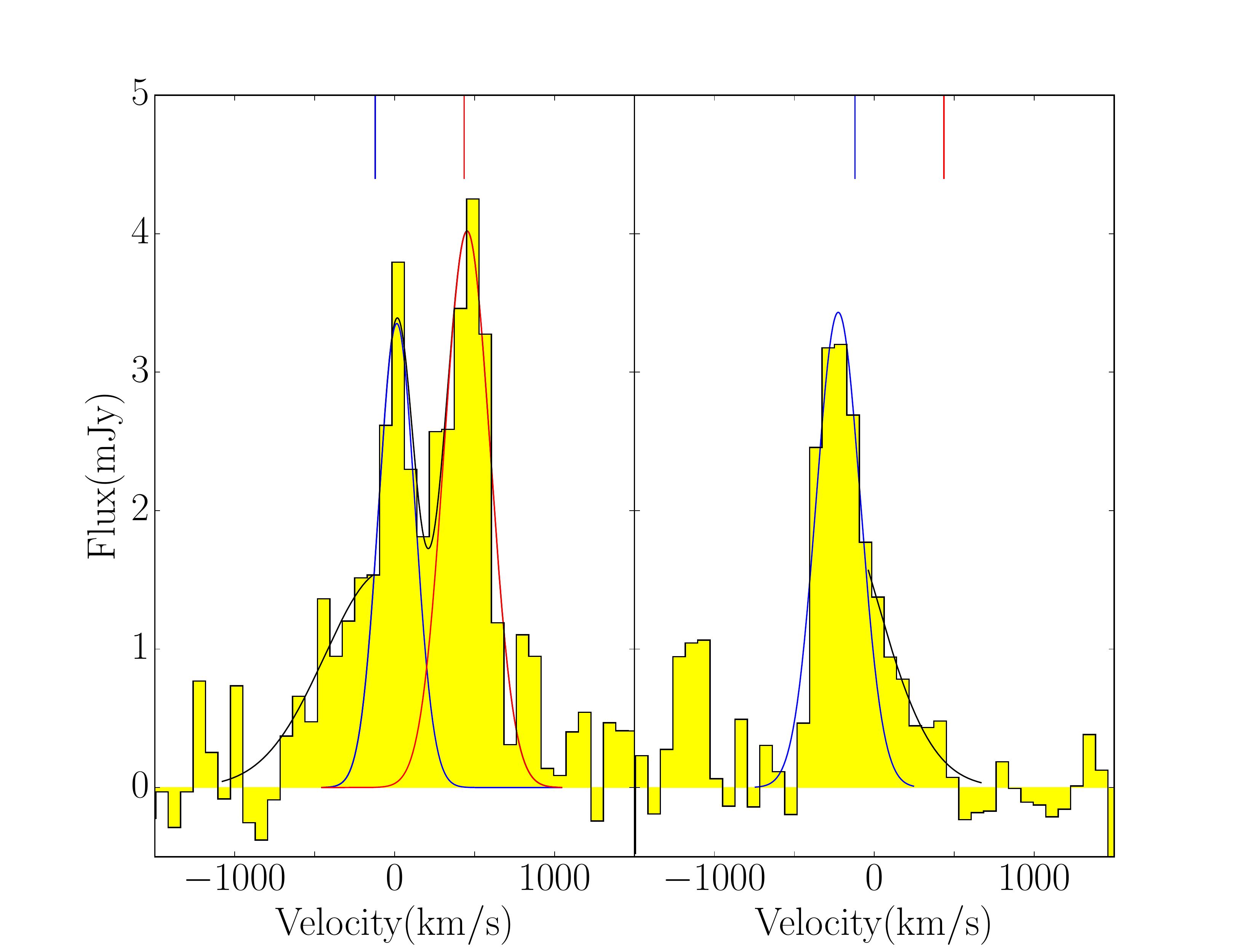}
\includegraphics[width=9cm]{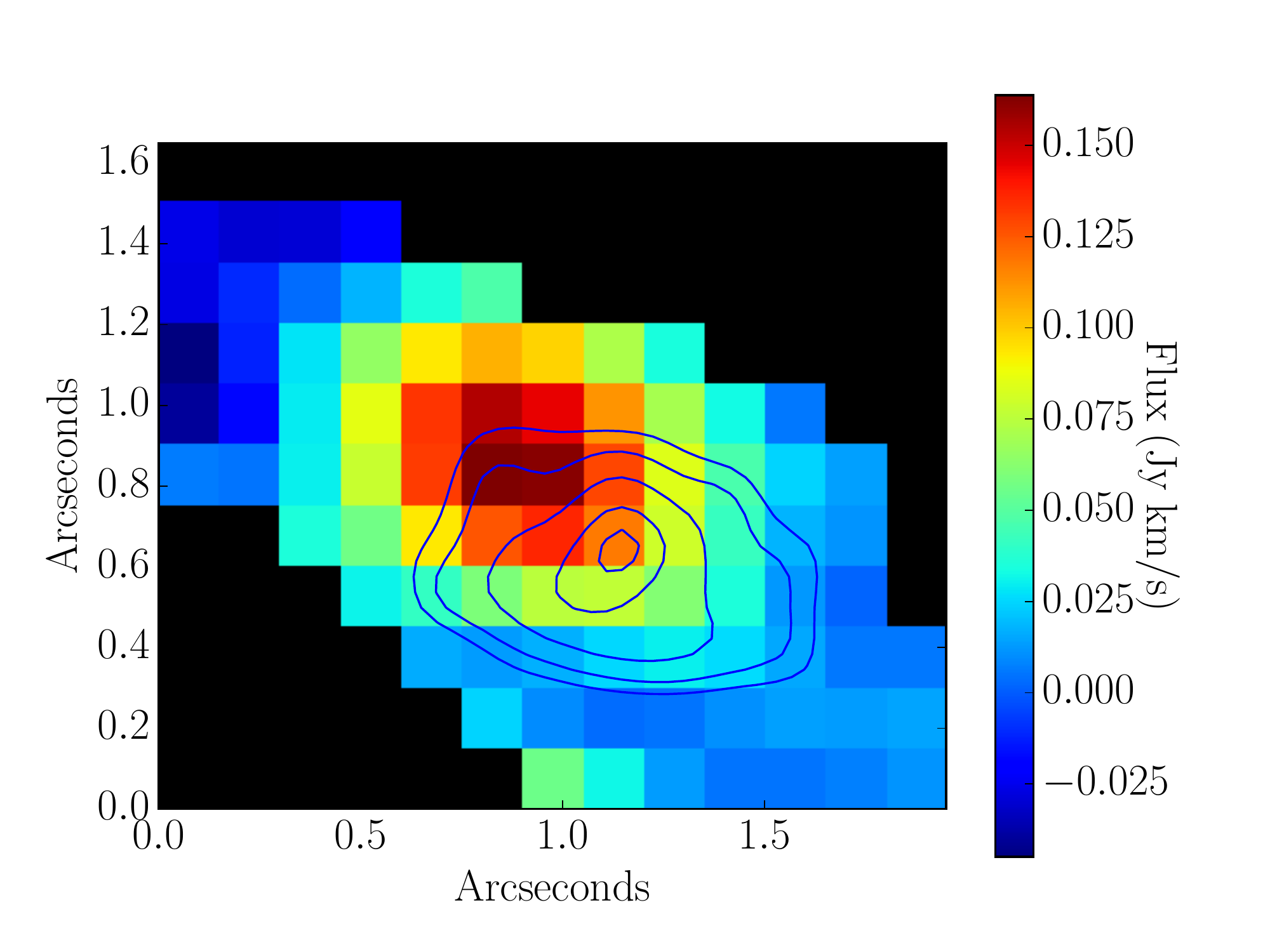}
\caption{A potential warp (or outflow) in the blue galaxy's gas disk -- {\bf Top Left: } A spectral region spanning $\sim$0.3$''$ found in the North-East section of the blue disk galaxy, with a clear wing on the blue side of the blue peak. The blue and red tick marks at the top of the plot correspond to the components systemic redshifts, respectively.   {\bf Top Right: } Similarly, a spectral region found in the South-West component of the blue source, with again, a wing on the side of the peak. The blue and red tick marks at the top of the plot correspond to the components systemic redshifts, respectively. 
 {\bf Bottom: } The integrated channel map of the wing (defined as $-$800 km/s to $-$450 km/s) that appears in the North-East region of the blue galaxy. The contours display the integrated channel map of the blue source, which shows a clear spatial offset with the wing and the blue galaxy by $0.5''$, well outside the full-width zero intensity regime of the blue line.}
 %It is more difficult to construct a similar channel map of the weaker positive velocity wing in the South-West, but exhibits a similar offset from the galaxy centroid. } %It would be impossible to display the blue galaxy's positive velocity wing here  in a similar manner, because of the overlapping red disk galaxy in velocity space.}
\end{figure}

\subsection{Potential wings to the CO lines}

The full line map of the merging complex (shown in Appendix-A) was also studied pixel by pixel for any features or irregularities in line profiles. 
Upon detailed inspection of the CO(5-4) spectra, the more massive (blue) component appears to show wings in the emission extending to velocities $\sim \pm1000$ km s$^{-1}$ from  systemic. 
The wings are only discernible in the off-nuclear regions, spatially offset by $\sim4.0$\ kpc in each direction along the major axis from the disk centroid.
An example of such spectra can be seen in Figure 4, which shows the highest significance spectra positions in   the North-East and South-West regions. 
The full spectral maps shown in Appendix-A  confirm visually that the central pixels do not exhibit any significant wings to the line. 
%Although it does seem that the wings extend in both the low and high velocities, the higher velocity line is difficult to quantify since the red emission line overwhelms any source of it. 

These wings are more prominent in the North-East region, and can be cleanly isolated here as a velocity channel map. 
Figure 4  shows the channel map for this low velocity wing,  defined between the range of $-$800kms to $-$450km/s, well outside the full-width zero intensity regime of the blue line. The blue components integrated channel map is contoured over the wing map, which clearly shows a spatial offset of 0.4$''$ from galaxy centroid to the extended line wing region.
It is more difficult to construct a similar channel map of the weaker positive velocity wing in the South-West, but it is clear from the emission line spectra that it exhibits a similar offset from the galaxy centroid. 
%
%It would be impossible to study the blue galaxy's positive velocity wing, if there were one,  in a similar manner because of the overlapping red disk galaxy in velocity space.
%However, clearly the wing offset to the South-West region is detectable and suggests the same may be true in the North-East.
%\textcolor{red}{is it strange that we say "its impossible to study the positive velocity wing" and then in the next line say "it is detectable though"?. Also, I'm not sure why we say "the same may be true in the north east" since we just explained that the more prominent wing is in the north east.}

These wings could be interpreted as  a warping of the gas disk from tidal forces, tidal features due to the merger \citep{2002MNRAS.333..481B} or even faint satellite galaxies around the blue component.
However wings on CO lines have also been interpreted as outflowing gas \citep{2012MNRAS.425L..66M}. 
The lack of wings on the CO line in the core region might be more consistent with such a {\it disk warp} interpretation. However, an episodic outflow driven by star formation or an Active Galactic Nucleus (AGN) could launch material out of the disk resulting in its detection only in these off-center locations (e.g., \citealt{2015A&A...574A..14C}).  %2015A&A...574A..14C 
%textcolor{red}{Ian: "So why do we conclude they are warps not outflows?"}
In \S~4.2 we assess these wings under these various possibilities.
%that they represent tidal forces warping the disk or outflowing gas.

\section{Discussion}

\subsection{Star formation in two massive disks}

We have demonstrated that \smghy\ is likely to represent two CO gas-rich disks undergoing a merger.
The merger itself is likely responsible for substantially elevating the far-infrared luminosity (and hence SFR) to  high levels. The very typical dust temperature (T$_{\rm d}$=36.1$\pm$3.1\,K) we have constrained for the system as a whole
%\textcolor{red}{(We dont state/constrain the total dust temperature anymore since we got rid of the SED fitting)}
suggests that a large dust masses is associated with the galaxies (via the large 850$\mu$m flux).
%The blue disk has a dust mass estimate approximately \textcolor{red}{3.8$\times$??} 2.5$\times$ larger than the red disk.
A gas mass ratio of the components of 1.4$\times$ is estimated from the Gaussian fits to the CO(3--2) peaks (\citealt{2015MNRAS.453..951C}).
With the majority of the 850$\mu$m flux attributed to the blue component (in a ratio potentially as high as $\sim$3$\times$), the dust mass of the blue component may be substantially higher than the red. However,  differences in dust temperature can affect this estimate  -- the difference in T$_{\rm d}$ would have to be about a factor of two to bring the dust mass ratio in line with the gas mass ratio between components (assuming two thirds of the 850$\mu$m emission was associated with the blue component).

Major mergers are  often invoked to explain very high SFR bursts, although there are counterexamples. %We have argued that the very stable disk properties (low Q values) suggest these two massive disk galaxies may be on a first approach, and are only beginning to merge.  SAID LATER
%Referring to the literature, observationally,  
\cite{2010ApJ...724..233E} %Engel et al.\ (2010) 
concluded, based on high-resolution CO imaging, that $z\sim2$ SMGs are typically nuclear starbursts on scales $<$ 4 kpc, driven by major gas-rich mergers, and hypothesize that mergers must be the driving mechanism behind the high SFRs.
However there have since been studies of SMGs that suggest that mergers are not necessary to trigger the starbursts, for instance 
\citet{2019MNRAS.490.4956G}.
%\textcolor{red}{Ian: "Not sure we believe this now."}
%
\cite{2008ApJ...684..996N} simulated a massive gas rich merger as the only plausible route to generating a SFR consistent with the S$_{850\mu \rm{m}}\sim$20\,mJy  SMG, GN20  (\citealt{2006MNRAS.370.1185P}). %Pope et al.\ 2006). 
However, high resolution followup of GN20 revealed a large undisturbed disk galaxy with no signs of mergers (\citealt{2012ApJ...760...11H}). %Hodge et al.\ 2012
%\textcolor{red}{More recent simulations of the most luminous SMGs have suggested very massive, isolated disk galaxies could exhibit large submm-wave luminosities comparable to this S(850$\mu$m)$\sim$20\,mJy regime (\citealt{2015Natur.525..496N}) IAN SAYS NOT CREDIBLE}.
%Narayanan D., et al., 2015
\cite{2010MNRAS.404.1355D} %Dave et al.\ (2010)
suggest that a substantial fraction of SMGs could be fed primarily by a cold mode accretion (e.g,. \citealt{2009Natur.457..451D} %Dekel et al.\ 2009a
; \citealt{2009MNRAS.395..160K}) % Keres  et al.\ 2009) 
and not major mergers.

%Recent hydrodynamical simulations of SMGs (Narayan et al.\ 2015) suggest that massive and turbulent disk galaxies can often show strong s and distortions without being in a state of ongoing merger or interation.

%Carilli et al.\ (2010) and Hodge et al.\ (2014) appeal to the large size of GN20 as argument against a major merger,

%%%
%In order to understand if our velocity spread made sense, a simulation was done to see if the radius we get for our galaxy is reasonable. Looking at figure 10, its clear that our radius of $\approx 6$kpc is within the simulation. 
%%%

Another question is whether  \smghy\ is  on its first approach, or has already gone through a close encounter.
%The CMA and major merger scenarios might be expected to have different structural and kinematic signatures. 
While relatively ordered rotation is observed in the extended gas distributions, the extended CO wings in the blue component (\S~3.4) suggest tidal forces at work.
%would favour  disks that are not being strongly disturbed. 
\smghy\ has a projected separation of 8\,kpc between the two massive disk components. 
The simulations in \cite{2015Natur.525..496N} %Narayan et al.\ (2015)  
suggest that the SFRs of the disks may not be  elevated by more than $\sim$20\% from tidal forces in the interaction if this is a first approach. 
%While the more massive of the two shows a clear warp in the gas disk, this is not unexpected for such a massive turbulent disk, and the warp is not clearly a sign that the merger is yet substantially  influencing the system or elevating the SFR. 
Further, comparison to these simulations suggests that our measured CO rotation out to $\sim$5\,kpc radius is not particularly unusual in these massive systems.
If we consider only the half-light size, from the exponential fit to the CO line intensity, the CO disk radius is  6\,kpc, in the upper third of massive systems considered in \cite{2015Natur.525..496N}. %Narayan et al.\ (2015)
The gas distributions and velocity fields (Figures 1\&2) of the disks in \smghy\ are also qualitatively similar to that seen in the simulated SMG maps of \cite{2010MNRAS.404.1355D} %Dave et al.\ (2010)
%particularly object ?B? which is a quiescently star- forming thick disk, 
though  the SFRs of these % z = 2 
simulated galaxies are substantially lower than \smghy.  \cite{2010ApJ...714.1407C} %Carilli et al.\ (2010) 
and \cite{2012ApJ...760...11H} %Hodge et al.\ (2012) 
used arguments of ordered disk-like gas motions to claim no major merger origin to the $z=4.05$ GN20 submm galaxy. However, such signatures do not conclusively rule out a merger, as \cite{2008ApJ...685L..27R} %Robertson \& Bullock (2008) 
have shown that ordered rotation of an extended gas disk can be reestablished very shortly after a major merger. 
It is possible that the \smghy\ component galaxies have already interacted with each other in a close encounter, but quickly reestablished their disk morphologies on a timescale shorter than the dynamical time $>$ 6kpc / 550 km s$^{-1}$ or $\simeq $11 Myr. 
% , we cannot make firm conclusions about the driver of star formation in GN20, 
%although the simplest interpretation of the molecular gas data favors CMA.

%These latter observations have now become possible with the JVLA in CO(1--0) and the NOEMA interferometer at its highest frequencies.

%\subsection{Simulation}

\begin{table}
 \flushleft{
  \caption{Derived properties of the individual red and blue components of \smghy. For the SFR, the estimates are based on the CO luminosity ratio applied to the total measured SFR from \S~3.4. 
  The integrated stellar mass was calculated in \citet{2015MNRAS.453..951C}  from a hyper-z fit to the IRAC and near-IR photometry.
  The final column presents either the sum of the two components (total), or the value derived from the combined or integrated system.
}
\begin{tabular}{llcccc}
\hline
 &  Red & Blue & Total/Combined  \\ 
%{}  & {(GHz)} &  {(hr)}  & (mJy) & {} \\ 
\hline
V$\mathrm{_{p-p}}$ ($\rm{km \ s^{-1}}$) & 196$\pm$13 & 303$\pm$15 & 854$\pm$42  \\
$\sigma$ ($\rm{km \ s^{-1}}$) & 136$\pm$33 &  180$\pm$39 & 375$\pm$54\\
Q & $1.20\pm0.18$& $1.50\pm0.20$& -- \\ %0.413 listed for total ?!?
SFR (M$_\odot \ \rm{s^{-1}}$)& \it{1600$\pm$300} & \it{2200$\pm$500} & 3800$\pm$600 \\
M$\mathrm{_{gas}}$ (M$_\odot$) & 7.0$\pm$1.1$\times10^{10}$ & 10.1$\pm$1.2$\times10^{10}$     & 17.5$\pm$1.4$\times10^{10}$ \\

\iffalse
M$_{dyn}$ (V) M$_\odot$.&  9.55$\times10^{9}$ & 2.40$\times10^{10}$ &  8.19$\times10^{10}$\\
M$_{dyn}$ (V$_{FWHM}$) M$_\odot$.&  1.55$\times10^{11}$ & 2.73$\times10^{11}$ & \\
Q (V) & 3.61 & 3.01&3.16\\
\fi
%M$\mathrm{_d}$ (M$_\odot$) & 1.1$\pm$0.3$\times10^{7}$ & 4.2$\pm$0.8$\times10^{7}$ & 5.3$\pm$1.0$\times10^{7}$\\
M$_*$ (M$_\odot$) & -- & -- & 5.8$\pm$1.5$\times10^{10}$ \\
M$\mathrm{_{dyn}}$ (M$_\odot$)  &0.71$\pm$0.22$\times10^{11}$  & 1.5$\pm$0.2$\times10^{11}$ &  
%\textcolor{red}
16 $\pm$ 2 $\times10^{11}$\\
%10.3$\pm$0.6$\times10^{11}$ \\
%I$_{\mathrm{wing}}$ (mJy km/s) & 0.253 & 0.703 & -- \\
\hline
\hline
\end{tabular}
} 
\end{table}

\subsection{Interpreting the CO line wings in the blue  component}
%A possible molecular outflow in the blue merger component}

%\begin{figure}
%\centering
%\includegraphics[width=9cm]{out_rate.pdf}
%\caption{Outflow mass-loss rate as a function of the SFR of \smghy.       Filled and unfiled circles represent unobscured and obscured AGNs respectively, LINERS are represented by triangles and pure starburst galaxies as stars (literature points taken from Ciccone et al.\ 2013). The black line represents the 1:1 correlation between outflows and SFR. HS1700.850.1 is shown as the green star, under assumption that the CO line wings represent outflowing gas. It lies near the 1:1 ratio, similar to other starbursts.}
%\end{figure}

%Due to the limited access of information available, more complex models cannot be investigated. Thus, 

The CO line wings in the \smghy\ blue component are intriguing. Given that a major merger has been characterized with a variety of evidence (\S~3.1), the wings may well be further evidence for the merger in the form of tidal warping of the gaseous disk. Most indicative, we see the line wings only at the edges of the disk (more susceptible to the torques from the merger) and not in the center.

\cite{2002MNRAS.333..481B} presented early evidence from simulations that mergers of massive disk galaxies can lead to strong and persistent tidal features in the gaseous and stellar components of the disks.
%Barnes (2002)
%
In a series of N-body experiments with merging massive galaxies, showing a range of pattern speeds, \cite{2009ApJ...703.2068D}, %Dubinski
found that the torques from the merger are strong enough to induce long-lived transient warps in the disks. These would appear similar to our CO wings found in \smghy, and with a similar luminosity ratio to the unwarped disk (with luminosity interpreted as molecular gas mass).
Even without a direct merger, strong warps are possible to the disks by nearby interactions. \cite{2014ApJ...789...90K} %Kim et al. (2014)
used N-body simulations to investigate the morphological and kinematical evolution of disks when galaxies undergo flyby interactions with adjacent dark matter halos, and found similar warping is possible.
With direct observations in the nearby LIRGs, NGC 3110 and NGC 232, \cite{2018ApJ...866...77E} %Espada et al. (2018),
showed very similar warping behaviour to the CO(2-1) lines from an SMA study of these merging disk galaxies.

A key question however is why other SMGs in this luminosity class do not typically exhibit the broad CO-line wings found in the blue component of \smghy? By comparison to the lower resolution data, it is clear that these faint wings only emerge through a careful high resolution analysis of the outer regions of the  velocity field of the blue disk, and is not discernable in the lower spatial resolution data (\citealt{2015MNRAS.453..951C}) or in the integrated line properties of the merging system.
It is thus possible that other SMGs studied in detail at high resolution may show similar features. 

Another possible interpretation of the broad wings on the CO lines is a molecular outflow. We view this as less likely as if they’re winds then we should see them across the disk, whereas no extended line profile is seen in the core of the galaxy. This would be possible if the winds were launched in the past and we are currently seeing them shut off \citep{2015A&A...574A..14C}, but is somewhat contrived given the immense ongoing starburst. 
For completeness we assess the basic properties of the emission line wings, assuming it is an outflow. Since we are unaware of the geometry of \smghy, we adopt a simple model following the arguments of \cite{2012MNRAS.425L..66M}, %Maiolino et. al (2012)
where the outflow occurs in a spherical volume with radius R$_{\mathrm{outfl}}$ which is uniformly filled with outflowing gas: %For this geometry, the outflow rate is given by
\begin{equation}
\dot{\mathrm{M}}_{\mathrm{outfl}} \approx \ v \  \Omega \ \mathrm{R^2_{outfl}} \ \langle{\rho_{\mathrm{outfl}}\rangle}\mathrm{v} = 3 \ v \ \mathrm{\frac{M_{outfl}}{R_{outfl}}}
\end{equation}
where $v$ is the outflow velocity, $\Omega$ is the total solid angle subtended by the outflow, and $\rho_{\mathrm{outfl}}$ is the volume averaged density of the gas in the outflow. %A lower limit on $\mathrm{M}_{\mathrm{outfl}}$ would then allow us to estimate a lower limit on the outflow rate. 
$\mathrm{M}_{\mathrm{outfl}}$ was calculated by determining the $L'$CO given its relationship to $\mathrm{M}_{\mathrm{gas}}$, assuming an $\alpha = 1$. (\citealt{2005ARA&A..43..677S}.) 
%Solomon and Vanden Bout, 2005.) 
An average CO flux (0.43 Jy km s$^{-1}$) was estimated with aperture photometry over the CO wing channels,
%,calculated by including all pixels in which the line wings were visible in our data, convolved with the b, giving a value of  0.43 Jy km s$^{-1}$. 
yielding a total mass of the outflow of 1$.99 \times10^{10} \ \rm{M}_{\odot}$. %A conservative value for 

Again following the arguments of \cite{2012MNRAS.425L..66M}, %Maiolino et. al (2012)
 the radius of the outflow is measured as 4.0\,kpc from the map in Figure 4, multiplying by 2 to account for matter ejected towards the outside of the galaxy. The velocity of the outflow is estimated to be the maximum value observed of 600 km $\mathrm{s^{-1}}$.
Given this, the outflow rate was calculated to be 
%was calculated by determining the distance between the maximum flux value for the blue galaxy and for the outflow, and then multiplying by 2 to account for matter ejected towards the outside of the galaxy. This gave a value of 1.8 kpc. 
%
%By using R$_{\mathrm{outfl}}$ = 1.8 kpc and v = 350 km s$^{-1}$ yields, 
%\begin{equation}
%\mathrm
${\dot{\rm{M}}_{\mathrm{outfl}}} > 4590 \ \mathrm{M_\odot \ yr^{-1}}$.
%\end{equation}
%
We also note  that if instead we assumed a shell like geometry, the outflow rate would be even higher.
The lower limit on the outflow rate is then %comparable to the
comparable to the 
SFR (2500 M$_\odot$ yr$^{-1}$) of the blue disk galaxy, %At the derived minimum outflow rate, 
and hence quenching star formation within less than 20 Myr (assuming a constant outflow rate with no new supply of gas from the outside.) %We note that if the outflow size has been overestimated, and the bulk of the gas is unresolved, then the lower limit on the outflow rate in Eq.\ 3 is even more conservative, since the outflow rate scales as $1/\mathrm{R_{outfl}}$. The kinetic power associated with the outflow is given by
%\begin{equation}
%\mathrm{P_K} \approx 0.5 \ \mathrm{v^2 \ \dot{M}_{outfl} } > 1.62 %\times10^{44} \ \mathrm{erg \ s^{-1}}
%\end{equation}
%
This interpretation of the wings on the CO lines provides an extreme outflow scenario compared to the literature, although consistent with expected scaling relations. 
The lack of any evidence for an AGN from X-ray and infrared SED \citep{2015MNRAS.453..951C} suggests that any outflow if real might be driven entirely by the copious star formation in the blue galaxy component. 
%While other interpretations of the velocity wings are likely possible, our analysis above shows that an outflow interpretation is not ruled out.

\subsection{The environment of \smghy\ }
%\textcolor{red}{Ian: "DELETE THIS SECTION?"}
Our deep (sub)mm-wave maps provide a route to searching for faint companions that may have gone unnoticed previously in submm studies of this field.
However no additional continuum sources were significantly detected in our combined SMA 870$\mu$m map reaching 0.9\,mJy rms (\S~2.2).
The combined A+D config 2mm continuum map from NOEMA is also substantially deeper than the D-config map presented in \cite{2015MNRAS.453..951C}. We searched the deconvolved map for additional point sources, but found none to a limit of 5$\sigma$ of 0.25\,mJy (\S~2.1).
We searched the full data cube over the 7000 km s$^{-1}$ probed by the WIDEX correlator band for any additional CO line candidates. We found no sources at $>5\sigma$ significance with kernels ranging from 200-1000 km s$^{-1}$. 

Thus aside from the on-going merger we have resolved with these new observations, there is no additional evidence for a locally overdense surrounding environment for \smghy.
As previously discussed, \smghy\ appears to reside in a relative void in the LBG redshift distribution of this field (\citealt{2015MNRAS.453..951C}), with no known companion UV-selected galaxies within $dz\sim0.05$ and 3$'$ (1.44 Mpc proper) radius.

In line with not living in an overdense environment, and by analogy with Hyperluminous quasars which have good statistical measurements of their environment (\citealt{2012ApJ...752...39T}), %Trainor \& Steidel 2012)
 the expected halo mass of \smghy\ should be no greater than $\sim10^{13}$ M$_\odot$.
Our dynamical measurement of the merging pair (\S3.2) puts a lower limit on the mass of
1.6 $\pm$ 0.2 $\times10^{12}$ M$_\odot$,
%\textcolor{red}{CAN'T REMEMBER WHERE WE GOT THIS VALUE???}, 
and comparison to other studies suggests this is also a typical mass for the observed  velocity separations of massive SMGs   (e.g. \citealt{2006MNRAS.371..465S}.)% Swinbank et al.\ 2006).

\subsection{The properties of the most luminous SMGs}

We next compare the properties of \smghy\ to other SMGs with comparably bright 850$\mu$m luminosities. 
%
%We have thus been able to substantially increase our detailed understanding of this hyper-luminous SMG. 
Sources such as \smghy\ remain very rare in surveys. ALMA and SMA followup to sources from the SCUBA-2 Cosmology Legacy Survey (S2CLS -- \citealt{2017MNRAS.465.1789G}%Geach et al.\ 2017
) presents the best benchmark for comparison. Only three sources with a single ALMA or SMA component S$_{850\mu m}>$19mJy exist in the $\sim$5deg$^2$ of the S2CLS, while only eight more such sources with S$_{850\mu m}>$15mJy are identified. 

Specifically, in the SMA followup of GOODS-N by 
\cite{2014ApJ...784....9B}%Barger et al.\ (2014)
, one source was found S$_{850\mu m}>$19mJy (the well studied GN20 -- \citealt{2012ApJ...760...11H}%Hodge et al.\ 2012
), and one was found S$_{850\mu m}>$15mJy. 
From the other northern S2CLS fields (AEGIS, SSA22, and Lockman),
\cite{2018MNRAS.477.2042H} %Hill et al.\ (2018) 
used the SMA in compact configuration to followup all the brightest 850$\mu$m sources $>8$mJy, and after deboosting, identified two SMGs with S$_{850\mu m}>$15mJy, both in the AEGIS field.

\cite{2018ApJ...860..161S} %Stach et al.\ (2018) 
did not find any sources with S$_{850\mu m}>$15mJy in the ALMA follow up of the 1.6\,deg$^2$ UDS field. There is a 30mJy source in the field, but it has been characterized as gravitationally lensed 
(\citealt{2015ApJ...810..133I}) %Ikarashi et al.\ 2015
and was not revisited in \cite{2018ApJ...860..161S}. %Stach et al.\ (2018)
\cite{2020MNRAS.495.3409S} %Simpson et al.\ (2020) 
used ALMA to follow up the 180 brightest sources in the COSMOS field (1.6\,deg$^2$), finding 10 with peak S$_{850\mu m}>$15mJy  and two with S$_{850\mu m}>$19mJy.
After deboosting, this becomes six sources with S$_{850\mu m}>$15mJy, only one of which has S$_{850\mu m}>$19mJy.
%in S2COSMOS there are 10 with peak S850>15mJy  (2 >19mJy) - but deboosting brings that down to 6 (with 2 > 19mJy).

Clearly these hyper-luminous submm sources show a great deal of cosmic variance. However, the statistics support our uncovering of \smghy\ in $\sim$0.3\,deg$^2$ of the KBSS followup as in line with these S2CLS surveys.
%While o
Our extended configuration SMA results suggest that the blue component of \smghy\ may carry the majority ($\sim$15\,mJy) of the 19.5mJy 870$\mu$m flux measured in the larger beam compact configuration observations.
%and a 4mJy source,
Our comparison sample could therefore be 11 sources in 5\,deg$^2$. However the SMA followup discussed above would not distinguish \smghy\ as two individual galaxies, and 3 in 5\,deg$^2$ remains a plausible statistic to compare.

Comparing some of the brightest SMGs that have been similarly characterized kinematically can provide some context to \smghy.
Another very bright source initially found from {\it Herschel} surveys (SGP38326 at  $z$=4.425, with S$_{850\mu m}$=23\,mJy -- 
\citealt{2016ApJ...827...34O}) %Oteo et al. 2016) 
is clearly characterized by a major merger, similar to \smghy,  supporting gas-rich mergers as a means to trigger such immense luminosities. \cite{2009MNRAS.400.1919N} %Narayanan et al.\ (2009) 
used hydro-dynamical simulations to highlight the final stage merger of two extremely gas rich, massive disk galaxies as a likely trigger of similarly intense starbursts.  

%The configurations of the mergers leading to intense starbursts can vary in the literature. 
Several studies have predicted that mergers configured with counter-rotating gas disks should lead to the most intense starbursts (e.g. \citealt{1994ApJ...431L...9M, 1996ApJ...464..641M}; %Mihos \& Hernquist 1994, 1996;
\citealt{1998ApJ...501L.167T}; %Taniguch\&Shioya 1998 
\citealt{2000ApJ...529L..77B}; %Borne et al. 2000;
\cite{2001ApJ...546..189B}; %Bekki 2001;
\citealt{2007A&A...468...61D}; %Di Matteo et al. 2007;
\citealt{2012A&A...545A..57S}%Salomé et al. 2012
). Although the components within SGP38326 (\citealt{2016ApJ...827...34O}) %Oteo et al.\ 2016
are resolved into counter-rotating gas disks, \smghy\ appears to resolve into two co-rotating disks, presenting a possible counterexample to the above studies. However we do not robustly constrain the inclination angle of the disks in \smghy, and they may still represent a merger quite far from co-rotating but seen in projection as roughly co-rotating.

%However, 
Other hyper-luminous SMGs do not show obvious signs of a major merger.
Of all the bright SMGs noted above in the S2CLS fields, only GN20 ($z$=4.055) has been studied at sufficiently high-resolution to compare to \smghy. High resolution followup in $^{12}$CO by \cite{2012ApJ...760...11H} %Hodge et al.\ (2012)
has demonstrated that GN20  exhibits highly symmetrical disk-like kinematics in its gaseous components. This is similar to another bright source found from {\it Herschel} surveys, SMM\,J084933 at $z$=2.410 (\citealt{2013ApJ...772..137I}). %Ivison et al. 2013 
While apparently not involved in a major merger, these bright SMGs both signpost large overdensities of galaxies, suggesting dense environments remain an important trigger of such luminous activity.
The gas disk in these cases may have recently reformed around the joint system.
Simulations in \cite{2015Natur.525..496N} %Narayanan et al.\ (2015)
 suggest that while the brightest SMGs are composed of multiple sources, %The simulation from Narayanan et al. 2015 shows that even  a major merger occurs, 
 the most active submm phase is often associated with a relatively isolated  period, several hundred Myrs after the most recent major merger. It is  possible  that GN20 and J084933 are consistent with this picture.

\section{Conclusions}
We have presented a high spatial resolution spectral analysis of the HyLIRG \smghy, the brightest 850$\mu$m source found in the SCUBA-2 KBSS fields. 

$\bullet$ We have resolved the CO emission into two components, which exhibit the expected properties of merging galaxies. This is supported by at least eight corroborating pieces of evidence (\S~3.1). Some of the stronger evidence includes the gap in the velocity field (which would otherwise require a central hole in the gas disk), and a range of differences in properties of the blue and red components. The $R_{53}$ CO line ratio  differences for instance would otherwise imply an inhomogeneous ISM within the system if it were a single large disk rather than two merging disks. Differing dust/gas masses, velocity gradients, and stability analysis within the two components are also difficult to explain in a single large rotating system.

$\bullet$ Each component of the merger shows well ordered gas motions of a massive, but turbulent disk. 
%\textcolor{red}{IAN: DELETE? 
The CO rotation curve of the blue disk shows some evidence for flattening of the rotation profile.
%where the dark halo may be dominating the mass profile.

$\bullet$ As we have resolved the size of each of the disks beyond the beam (0.8$''$ beam size along the rotation axis), the 0.4$''$ FWHM of the beam minor axis puts a strong constraint on the inclination of each of the disks, and removes some of the $\sin(i)$ uncertainty in the mass estimates from rotation. 
%$\bullet$
We estimate disk masses of the blue and red components, finding that they are of order 10$^{11}$\,M$_\odot$. %with up to a factor two difference between the two components depending on kinematic modelling.

$\bullet$ The more massive (blue) component shows extended wings in CO(5--4) due to either a warp/tidal feature, or less likely a molecular outflow in the gas disk. %\textcolor{red}{(DELETE) at the outer edges}. %the first ever found in a distant SMG. 
%The lack of any evidence for an AGN (neither in 200\,ksec {\it Chandra} X-ray, nor the infrared SED) suggests that if this is an outflow it would need to be driven entirely by the copious star formation in the blue galaxy component.

A number of key observations are required to untangle the mechanisms driving star formation in \smghy. Higher-resolution imaging of the millimeter continuum would better reveal the distribution of star formation across the disk in greater detail. While our NOEMA and SMA continuum measurements have resolved the two merging disks separated by only 0.7$''$, the beamsizes are comparable to the separation. 
%we have as of yet no measurements of the resolved continuum of each disk.
Higher-resolution imaging of low-$J$ and high-$J$ CO transitions can be used to determine the spatial excitation of the molecular gas, which will be possible with recent and upcoming baseline increases at NOEMA. Our observations have succeeded in resolving the gas beyond the beam in each disk,  showing clear rotation patterns. %\textcolor{red}{GET RID OF? possibly flattening from the massive dark matter halos}. 
However the internal excitation of the gas in the disks have yet to be determined, requiring more than one CO transition. %And third, high  spatial resolution imaging of the low-excitation CO emission is required to determine the gas dynamics on small scales.
%i.e., verify the overall rotation, and determine the internal turbulent velocity and stability parameters in the disk. 
\cite{2008ApJ...682..231S} %Shapiro et al.\ (2008) 
show that with such resolved measurements, the higher order  {\it kinemetry} of the line maps can further enable an empirical differentiation between merging and non-merging galaxies.

We have thus been able to substantially increase our detailed understanding of this hyper-luminous SMG. Such sources remain very rare -- only three comparably bright S$_{850\mu m}>$\,19mJy sources exist in the $\sim$5deg$^2$ of the S2CLS Legacy survey, followed up with ALMA and SMA, while only eight more sources with S$_{850\mu m}>$15\,mJy are identified.
Comparing some of the brightest SMGs that have been similarly characterized kinematically, we find they often show explicit signs for being in a  merger, or else are signposting a massive overdensity of galaxies, indicating that recent mergers were likely and abundant gas supplies feeding the SMG are likely present.
\smghy, despite being in an active merger, remains somewhat unique in the literature of  hyper-luminous SMGs, in residing in such a rarefied environment lacking submm companions or an  overdensity of galaxies at the same redshift.

%While our results suggest that we have resolved \smghy\ into a 15.7mJy and a 4mJy source, and thus our comparison sample is 10 sources in 

\section*{Acknowledgements}

This work is based on observations carried out under project number S17BS with the IRAM NOEMA Interferometer. IRAM is supported by INSU/CNRS (France), MPG (Germany) and IGN (Spain). %This research made use of Astropy, a community-developed core Python package for Astron- omy (the Astropy collaboration 2013). %Support for RD was provided by the DFG priority program 1573 ÒThe physics of the interstellar mediumÓ. 
The Submillimeter Array is a joint project between the  Smithsonian Astrophysical Observatory and the Academia Sinica Institute of Astronomy and Astrophysics and is funded by the Smithsonian Institution and the Academia Sinica.
SC acknowledges support from NSERC and CFI.
IRS acknowledges support from STFC (ST/T000244/1).

\section*{Data Availability Statement}
The data underlying this article will be shared on reasonable request to the author.

%%%%%%%%%%%%%%%%%%%%%%%%%%%%%%%%%%%%%%%%%%%%%%%%%%

%%%%%%%%%%%%%%%%%%%% REFERENCES %%%%%%%%%%%%%%%%%%

%The best way to enter references is to use BibTeX:

\bibliographystyle{mnras}
%\nocite{*}
\bibliography{bib_test} 
% if your bibtex file is called example.bib

% Alternatively you could enter them by hand, like this:
% This method is tedious and prone to error if you have lots of references
%\begin{thebibliography}{99}
%\bibitem[\protect\citeauthoryear{Author}{2012}]{Author2012}
%Author A.~N., 2013, Journal of Improbable Astronomy, 1, 1
%\bibitem[\protect\citeauthoryear{Others}{2013}]{Others2013}
%Others S., 2012, Journal of Interesting Stuff, 17, 198
%\end{thebibliography}

%%%%%%%%%%%%%%%%%%%%%%%%%%%%%%%%%%%%%%%%%%%%%%%%%%

%%%%%%%%%%%%%%%%% APPENDICES %%%%%%%%%%%%%%%%%%%%%

\appendix

\section{Full spectral map of \smghy}
In this appendix, the full mosaic of CO emission line spectra are shown for each 0.13$''$ pixel in the \smghy\ system. 
%Pixels are shown down to a CO(5--4) line detection limit of 3$\sigma$. 
The line fitting code used to make the moment maps is described in Appendix~B.

%In this appendix, we show the results of our line fitting code developed to analyze the \smghy\ system. Lines are fit down to a detection limit of 3$\sigma$, and a prior on criteria for fitting a given pixel is imposed based on the adjacent pixels (the velocity cannot shift by more than 50 km s$^{-1}$, and the linewidth cannot change by more than 50 km s$^{-1}$).

\begin{figure*}{}
\includegraphics[width=22cm,  angle =270 ]{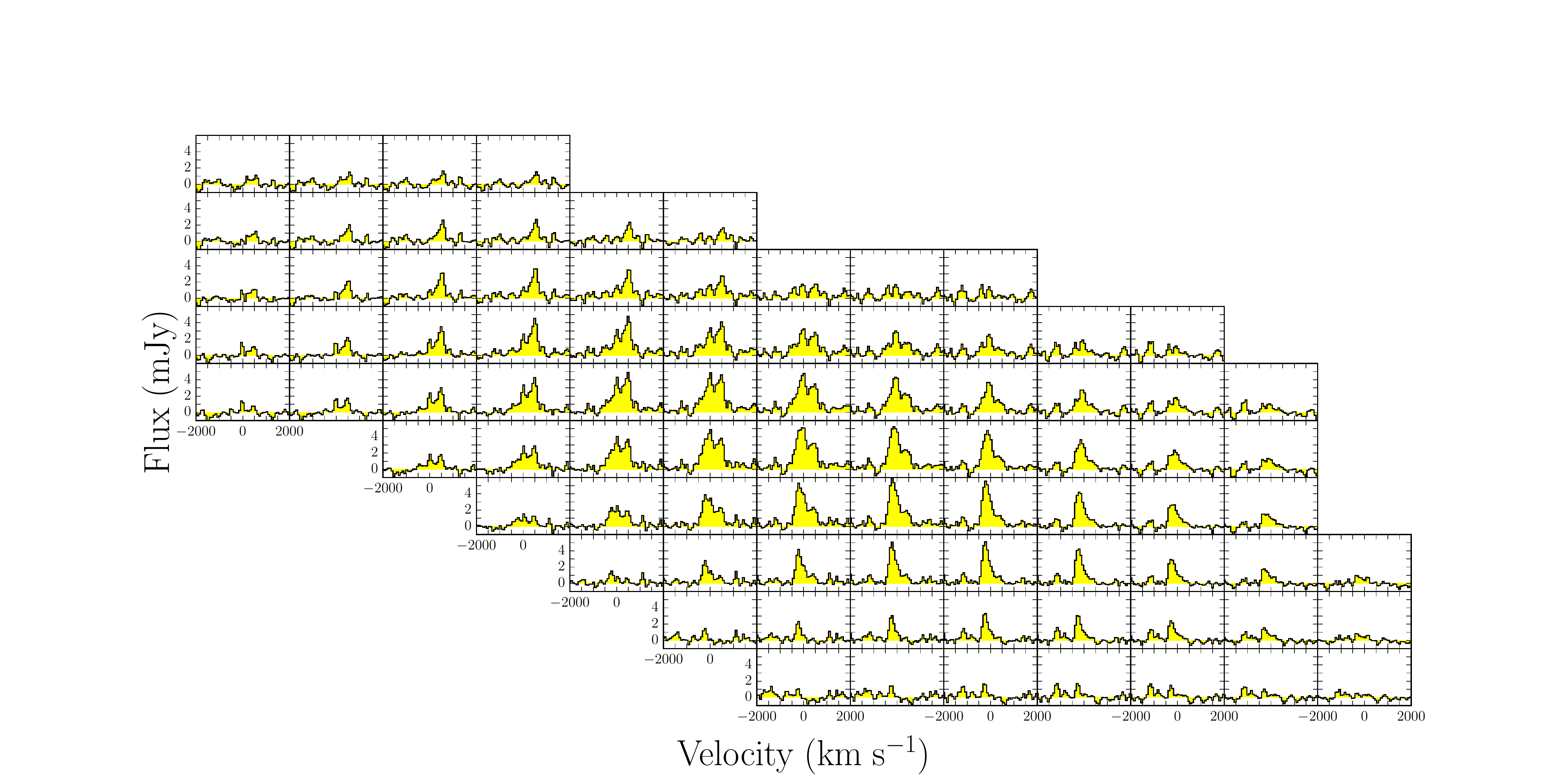}
%\vskip-2cm 
\caption{The full mosaic of 1D spectral extractions for the \smghy system. 
%extending to the 3$\sigma$ line detection limit in each direction.
Each box represents a pixel of 0.13$''$ (units of mJy per beam in each pixel), and the measurements extend over 1.7$''$ in RA and 1.3$''$ in DEC.  The synthesized beam size is 0.8$''$ $\times$ 0.4$''$ (\S~2.1) along a position angle 59 deg East of North. }
\end{figure*}

\clearpage 
%\onecolumn
\section{Line Fitting / Moment Map Python Code}
Emission line fitting is a vital component for analyzing the kinematics of galaxies. By fitting Gaussian emission lines to the spectra, it allows us to determine the average velocity of the gas, integrated flux and the velocity dispersion of the galaxies at a specific location. Putting all these together, we can look at the total kinematics of the sources as a whole. The code used for the emission line fitting throughout the paper is compatible with python version 2.7. Below are the key points of the code and an explanation of how it works:

\begin{itemize}
  \item The code initially extracts the data from the data cube, specifying the $x$ and $y$ pixel coordinates and the emission spectra for that specific pixel
  \item A radius is chosen for which the code looks to fit a Gaussian to the emission spectra, which it then loops over and discards any pixel with an integrated flux of < $3\sigma$ 
  \item The Gaussian profiles are fit using the scipy.optimize.curve\textunderscore fit function, which uses a nonlinear least squares fitting method to determine the best fit to the given data with a specified model. 
  \item The first function to be fit to the spectra is a Gaussian plus continuum level, which is then subtracted off the total spectra
  \item The gaussian function is forced to try fitting at the maximum value of the spectra (i.e. the peak of the emission line and the average velocity of the gas) so it fits the correct portion of data. The dispersion is then determined by the fitting parameters calculated. The errors for each parameter are calculated through the scipy.optimize.curve\textunderscore fit function. 
  \item Finally, the integrated velocity flux is calculated by integrating over the total emission line model once it has been fit
  \item Once this process has been completed for every pixel in the radius specified, moment maps are created by placing the values of each parameter in the corresponding pixel location of the galaxy
\end{itemize}

A sample code will be made available online.

\bsp	% typesetting comment
\label{lastpage}
\end{document}